\documentclass[apj, twocolumn]{aastex63}

\usepackage{natbib}
\usepackage{hyperref}

\usepackage{booktabs}

\usepackage[T1]{fontenc}
\usepackage{graphicx}
\usepackage{bm}

\hypersetup{colorlinks=true, citecolor=blue, linkcolor=cyan, urlcolor=magenta}

\usepackage{epstopdf} 
\usepackage{lipsum}

\newcommand{\HI}{\ion{H}{1}}
\newcommand{\alf}{$\alpha.40$}

\begin{document}

\title{Connecting Optical Morphology, Environment, and \HI{} Mass Fraction for Low-Redshift Galaxies Using Deep Learning}

\author[0000-0002-5077-881X]{John F. Wu}
\affiliation{Department of Physics \& Astronomy, Johns Hopkins University, 3400 N. Charles Street, Baltimore, MD 21218, USA}
\affiliation{Space Telescope Science Institute, 3700 San Martin Drive, Baltimore, MD 21218, USA}
\email{jfwu@jhu.edu}


\begin{abstract}
A galaxy's morphological features encode details about its gas content, star formation history, and feedback processes, which play important roles in regulating its growth and evolution.
We use deep convolutional neural networks (CNNs) to learn a galaxy's optical morphological information in order to estimate its neutral atomic hydrogen (\HI{}) content directly from SDSS $gri$ image cutouts.
We are able to accurately predict a galaxy's logarithmic \HI{} mass fraction, $\mathcal M \equiv \log(M_{\rm HI}/M_\star)$, by training a CNN on galaxies in the ALFALFA 40\% sample.
Using pattern recognition (PR), we remove galaxies with unreliable $\mathcal M$ estimates.
We test CNN predictions on the ALFALFA 100\%, xGASS, and NIBLES catalogs, and find that the CNN consistently outperforms previous estimators.
The \HI{}-morphology connection learned by the CNN appears to be constant in low- to intermediate-density galaxy environments, but it breaks down in the highest-density environments. 
We also use a visualization algorithm, Gradient-weighted Class Activation Maps (Grad-CAM), to determine which morphological features are associated with low or high gas content. 
These results demonstrate that CNNs are powerful tools for understanding the connections between optical morphology and other properties, as well as for probing other variables, in a quantitative and interpretable manner.
\end{abstract}

\section{Introduction}

Neutral atomic hydrogen (\HI{}) is the dominant component of cool gas in the interstellar medium (ISM) for low-redshift galaxies \citep[e.g.,][]{Saintonge+17}.
Although neutral gas is crucial for understanding how galaxies evolve and grow over cosmic timescales, \HI{} is difficult to detect in extragalactic sources because of its weak 21-cm emission line.  
This observational challenge has been partially mitigated by large \HI{} surveys such as the \HI{} Parkes All Sky Survey \citep[HIPASS;][]{Barnes+01}, the Arecibo Legacy Fast ALFA Survey \citep[ALFALFA;][]{Giovanelli+05}, and the GALEX Arecibo SDSS Survey \citep[GASS;][]{Catinella+10}, which have taken a census of the brightest \HI{} sources in the local Universe.
New radio telescopes such as MeerKAT, ASKAP (Australian Square Kilometre Array Pathfinder), and eventually the SKA will allow us to measure \HI{} at much lower masses ($M_{\rm HI}$) and at higher redshifts; see, e.g., Looking at the Distant Universe with the MeerKAT Array \citep[LADUMA;][]{Blyth+16}, MeerKAT International GHz Tiered Extragalactic Exploration \citep[MIGHTEE;][]{Jarvis+16}, Wide-field ASKAP L-Band Legacy All-sky Blind surveY \citep[WALLABY;][]{WALLABY}, and Deep Investigation of Neutral Gas Origins (DINGO).\footnote{\url{https://dingo-survey.org/}}

Small and incomplete \HI{} samples currently limit our ability to study gas properties in typical galaxies beyond $z \approx 0.05$.
Since \HI{} is so important to galaxy evolution but challenging to measure, astronomers have devised proxies for estimating galaxies' gas content.
For example, \cite{Kannappan04} proposed ``photometric'' gas fractions, leveraging the valuable connection between global gas content and optical properties.
\cite{Zhang+09} tightened the relationship by accounting for $i$-band surface brightness in addition to $g-r$ color.
More complicated heuristics and machine learning models have also been used \citep[e.g.,][]{Teimoorinia+17,Rafieferantsoa+18}, although these estimators become more difficult to interpret as the number of parameters increases.
Indeed, computer vision algorithms seem to perform spectacularly well at predicting galaxy properties directly from optical imaging \citep[e.g.,][]{Dieleman+15,Huertas-Company+19,Morningstar+19,Pasquet+19,WuBoada19}, but because these models often have millions of parameters, it can be difficult to understand what makes them so successful.

We train a deep convolutional neural network (CNN) to directly predict the logarithm of the \HI{} mass fraction, $\mathcal M \equiv \log (M_{\rm HI}/M_{\star})$, using three-band optical image cutouts from the Sloan Digital Sky Survey (SDSS).
After demonstrating that our trained model can predict $\mathcal M$ to within 0.23~dex for \alf{}, we test the CNN on independent data sets and examine which factors result in poor predictions. 
We use the same CNN method to distinguish ALFALFA detections from non-detections, and estimate a galaxy's likelihood of detection in an ALFALFA-like survey from its $gri$ imaging.
Using this pattern recognition system, we evaluate only the robust predictions on test data sets again and compare to results in the literature.
We investigate how the relationship between optical imaging and \HI{} content varies with galaxy environment.
We also use the Grad-CAM algorithm to localize image features that the CNN associates with high or low gas mass fraction in order to visually interpret which morphological features are relevant to machine learning predictions; it essentially tells us which parts of the image the CNN is looking at in order to determine the gas mass fraction \citep[see, e.g.,][]{PeekBurkhart19}.

The paper is structured as follows.
We describe the \HI{} catalogs and optical imaging in Section~\ref{sec:data}, and explain some details of the CNNs in Section~\ref{sec:CNNs}.
In Section~\ref{sec:results}, we present our results showing that a CNN trained on ALFALFA can accurately recover $\mathcal M$, and report test results on independent data sets.
In Section~\ref{sec:PR}, we use pattern recognition to identify galaxies that are expected to be detected by an ALFALFA-like survey, and in Section~\ref{sec:T17-comparison} we compare our results to other $\mathcal M$ estimators in the literature.
In Section~\ref{sec:environment}, we quantify the impact of environmental effects and study how the connection between \HI{} content and optical morphology breaks down in the most overdense environments.
In Section~\ref{sec:interpretation}, we discuss and interpret the morphological features that a CNN ``sees'' in order to distinguish gas-rich systems from gas-poor galaxies.
We discuss future prospects in Section~\ref{sec:future}, and report our conclusions in Section~\ref{sec:conclusions}.
In the Appendix, we present comparisons between CNNs and simpler machine learning models and tests of CNN performance when artificial sources are added to the input images.
Throughout this paper, we assume a cosmology with $H_0 = 70~{\rm km~s^{-1}~Mpc^{-1}}$, $\Omega_m = 0.3$, $\Omega_\Lambda = 0.7$. 
All of the code used in our analysis is publicly available at \url{https://github.com/jwuphysics/HI-convnets}.

\section{Data} \label{sec:data}

\begin{figure}[t!]
    \includegraphics[width=\columnwidth]{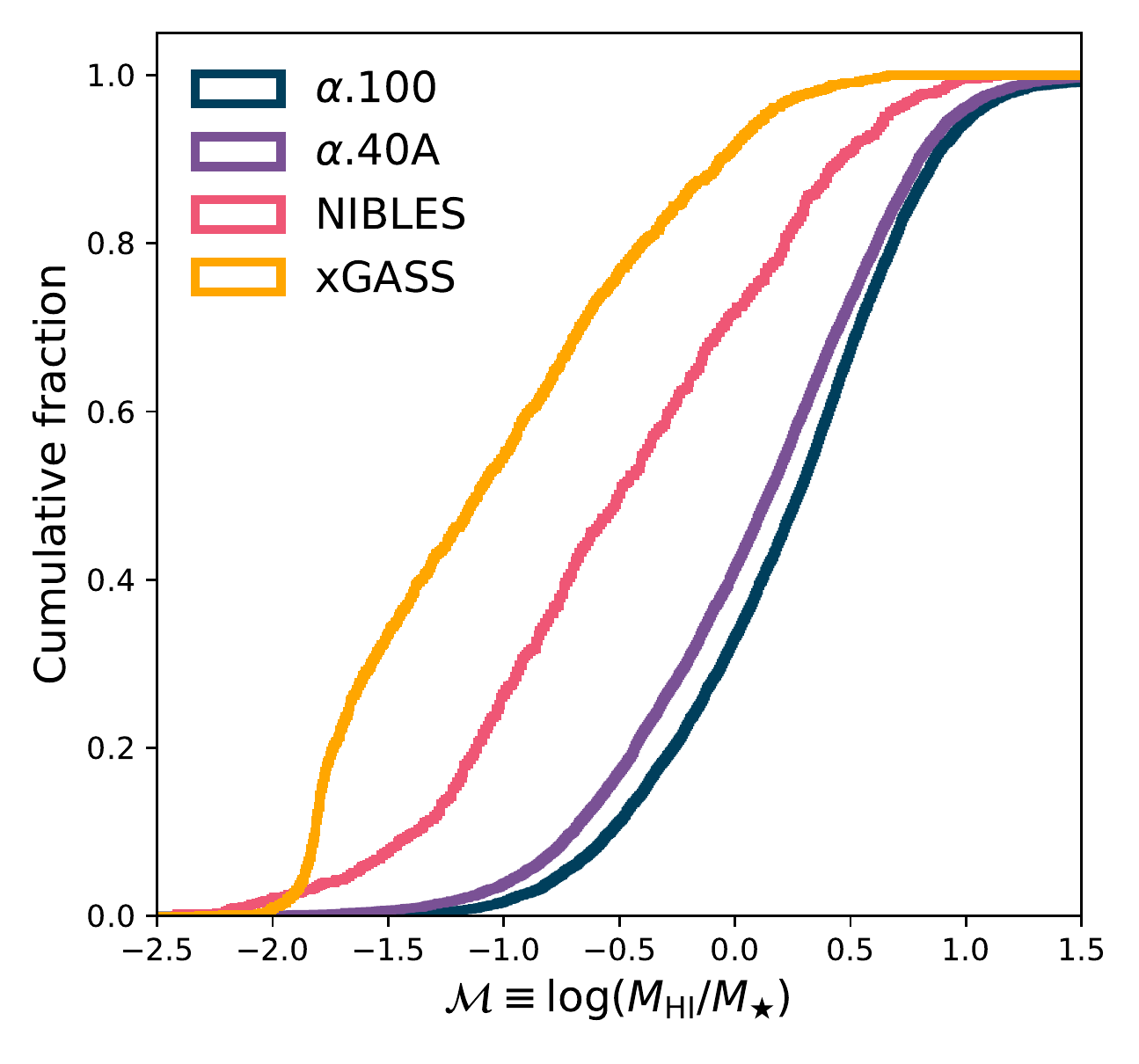}
    \caption{\label{fig:M-distribution}
    \HI{} mass fraction cumulative distribution functions for the ALFALFA parent samples, xGASS representative sample, and NIBLES SDSS-crossmatched sample.
}
\end{figure}

We make use of four \HI{} data sets in our analysis: \alf{}, $\alpha$.100, NIBLES, and the xGASS representative sample.
Because these data sets have different selection criteria, they are useful for testing our CNN methods on galaxy samples with varying \HI{} mass fraction distributions.
In Figure~\ref{fig:M-distribution}, we show the cumulative distribution functions of $\mathcal{M}$, the logarithmic \HI{} mass fraction.
The catalogs' stellar and \HI{} properties are also summarized in Table~\ref{tab:summary}.
We describe the data sets and our selection criteria in greater detail below.

\paragraph{SDSS imaging} 
The ALFALFA, NIBLES, and xGASS data sets have publically available catalogs of optical counterparts.
We obtain $gri$ imaging from the SDSS DR14 \citep{Abolfathi+18} SkyServer using the Image Cutout service\footnote{\url{http://skyserver.sdss.org/dr14/en/help/docs/api.aspx}} queried via a custom Python script.
The conversion of $gri$ imaging to RGB channels is a modified version of the \cite{Lupton+04} algorithm, as described on the SkyServer website.\footnote{\url{https://www.sdss.org/dr14/imaging/jpg-images-on-skyserver/}}
Downloaded JPG images have $224 \times 224$ pixels at the native SDSS pixel scale (0.396\arcsec~pixel$^{-1}$),
which corresponds to angular sizes of $1.48\arcmin \times 1.48 \arcmin$. 

\paragraph{Stellar masses} 
In order to compute gas mass fractions, \HI{} detections are crossmatched with galaxies in the SDSS DR7 MPA-JHU value-added catalog \citep{Kauffmann+03,Brinchmann+04,Tremonti+04,Salim+07}.
All stellar mass estimates assume a \cite{Chabrier03} initial mass function.
ALFALFA is crossmatched on the basis of \texttt{PhotoObjID} (\alf{}) or a 1\arcsec{} search radius ($\alpha$.100).
The xGASS and NIBLES catalogs already include crossmatched stellar masses.
We keep only the galaxies with valid stellar mass estimates. 

\begin{deluxetable*}{l r RRR c RRR c RRR}
    \tablewidth{0pt}
    \tablecolumns{13}
    \tablecaption{\HI{} data sets \label{tab:summary}}
    \tablehead{
        \colhead{Data set\tablenotemark{a}} & 
        \colhead{$N$} &
        \multicolumn{3}{c}{$\log(M_{\rm HI}/M_\sun)$} & \colhead{} &
        \multicolumn{3}{c}{$\log(M_{\star}/M_\sun)$} &
        \colhead{} &
        \multicolumn{3}{c}{$\mathcal M_{\rm true}$} \\
        \cline{3-5} \cline{7-9} \cline{11-13}
        \colhead{} &
        \colhead{} &
        \colhead{0.16} & \colhead{0.50} & \colhead{0.84} &&
        \colhead{0.16} & \colhead{0.50} & \colhead{0.84} &&
        \colhead{0.16} & \colhead{0.50} & \colhead{0.84}
    }
    \startdata
    \alf{}A & 7128 & 9.21 & 9.71 & 10.06 && 8.71 & 9.56 & 10.41 && -0.52 & 0.15 & 0.68 \\
    \alf{}B & 4644 & 9.21 & 9.68 & 10.02 && 8.73 & 9.41 & 10.07 && -0.26 & 0.25 & 0.70 \\
    $\alpha$.100 & 6087 & 9.22 & 9.67 & 10.03 && 8.62 & 9.38 & 10.24 && -0.37 & 0.28 & 0.75 \\
    NIBLES & 899 & 8.54 & 9.18 & 9.73 && 8.52 & 9.62 & 10.68 && -1.19 & -0.50 & 0.29\\
    xGASS & 1179 & 8.68 & 9.27 & 9.84 && 9.52 & 10.30 & 10.95 && -1.78\tablenotemark{b} & -1.11 & -0.27 \\
    \enddata
    \tablecomments{
    For each data set, we show the number of galaxies ($N$) after crossmatching to SDSS, performing all cuts, and removing sources in common with other catalogs.
    We report 0.16, 0.50 (median), and 0.84 quantile values for the \HI{} mass, stellar mass, and gas mass fraction.
    }
    \tablenotetext{a}{We have removed overlaps between the \alf{}A, $\alpha$.100, NIBLES, and xGASS samples. \alf{}B is a subset of \alf{}A.}
    \tablenotetext{b}{The xGASS catalog includes upper limits on $\mathcal M_{\rm true}$ (see text for details).}
\end{deluxetable*}

\subsection{ALFALFA \alf{}} \label{sec:a40}
The Arecibo Legacy Fast ALFA (ALFALFA) survey is a $z \leq 0.06$ blind search for \HI{} at high Galactic latitudes \citep{Giovanelli+05}.
The ALFALFA \alf{} catalog covers 40\% (2800~deg$^2$) of the full survey area \citep{Haynes+11}; most of these detections (12,468 sources) lie within the SDSS footprint.
Our sample includes rare, high-mass \HI{} systems that are not necessarily representative of the probed cosmic volume.
There also exists a nearly volume-limited ALFALFA subsample at $z \leq 0.05$, but we are interested in training our CNN with the larger data set.
We select sources with \texttt{OCcode = I} in order to retain \alf{} detections with SDSS counterparts, and we drop all sources with duplicate matches to DR7 identifiers.
This cut reduces the number of \HI{} sources to 11,739.
We create the \alf{}A catalog, which contains 7,399 galaxies with valid stellar mass estimates, and the \alf{}B catalog, a subset of \alf{}A containing 4,797 galaxies with valid $M_\star$, SFR, gas metallicity, and spectroscopic redshift measurements.

\subsection{ALFALFA $\alpha$.100}\label{sec:a100}

The $\alpha$.100 catalog contains 31,502 extragalactic sources across the full ALFALFA sky \citep{Haynes+18}.
A large fraction of these sources do not overlap with the SDSS footprint, and a significant subset of $\alpha$.100 is already catalogued in \alf{}.
Since there is no public combined catalog for $\alpha$.100 with SDSS identifiers, we perform a custom crossmatching exercise. 
We crossmatch sources in the ALFALFA Spring sky to the SDSS MPA-JHU catalog using a 1\arcsec{} radius, exclude systems with other \HI{} counterparts within a 1.9\arcmin{} radius (to avoid ALFALFA source blending), and exclude SDSS galaxies with neighboring optical sources within a 55\arcsec{} radius (to avoid fiber collisions).
Since we will primarily be using $\alpha$.100 as an independent test data set, we also remove duplicates of \alf{}A and xGASS sources in $\alpha$.100 via another positional crossmatch (with search radius of 1\arcsec{}).
These selection criteria leave 6,087 galaxies in the $\alpha$.100 catalog.
Among the \HI{} catalogs, the $\alpha.100$ sample has the highest gas mass fractions (Figure~\ref{fig:M-distribution}).

\subsection{NIBLES} \label{sec:NIBLES}

The Nan\c{c}ay Interstellar Baryons Legacy Extragalactic Survey (NIBLES) catalog of \HI{} detections contains 1,864 low-redshift galaxies with heterogeneous absolute $z$-band magnitudes \citep{NIBLES}.
The NIBLES sample is characterized by a wide range of stellar masses and intermediate values of $\mathcal{M}$ relative to \alf{} and xGASS (Table~\ref{tab:summary}).
We remove systems with very low ($< 10^8~M_\sun$) or unavailable MPA-JHU stellar mass estimates.
We visually inspect the SDSS image cutouts and eliminate sources with no apparent optical counterpart near the center, those with only a point source in the center, and those significantly corrupted by imaging artifacts, leaving 941 galaxies.
Finally, we also remove sources that overlap with ALFALFA and/or xGASS.
There are 899 remaining galaxies in the cleaned NIBLES data set.

\subsection{xGASS representative sample} \label{sec:xGASS}
ALFALFA detections tend to be the most \HI{}-rich systems in the local Universe, and differ from the majority of galaxies found in optical surveys.
We use the extended GALEX Arecibo SDSS Survey representative sample  \citep[xGASS;][]{Catinella+18} in order to repeat our analysis on galaxies with more typical star formation and gas properties.
xGASS consists of 1,179 galaxies with stellar masses between $9 \leq \log (M_\star / M_\sun) \leq 11.5$ in the redshift range $0.01 \leq z \leq 0.05$.
It is evident from Figure~\ref{fig:M-distribution} that xGASS galaxies are the most gas-poor of our three samples, in part due to their relatively high stellar masses (Table~\ref{tab:summary}).
All xGASS systems have ancillary SDSS photometry and spectroscopy.
The sample spans a range of galaxy morphologies, from passive ellipticals to starbursting mergers, and is complete down to $\mathcal M \approx -1.70$ for galaxies with $\log(M_\star / M_\sun) \geq 9.7$.
The most gas-poor members of the xGASS sample only have $5\,\sigma$ upper limits on $M_{\rm HI}$ available.
However, we include them in our sample because the more massive systems have been observed to similar $\mathcal M$ completeness, and the entire sample has a common gas mass limit $\log(M_{\rm HI}/M_\sun) = 8$.

\section{Methodology: deep neural networks} \label{sec:CNNs}

We optimize deep CNNs in order to predict the \HI{} mass fraction directly from three-band SDSS images, i.e., arrays of $3 \times 224 \times 224$ pixels.
Because the goal is to estimate $\mathcal M$, a scalar quantity, we are optimizing the CNN to solve a regression task rather than a classification problem (although in later sections we will also use CNNs for classification).
Training a neural network requires several steps, which can briefly described as follows.
The CNN ingests a batch of images and outputs predictions ($\mathcal M_{\rm pred}$) one batch at a time.
These predictions are compared to their ground truth values (i.e., $\mathcal M_{\rm true}$) via the \textit{loss function}, which measures the level of discrepancy.
CNN model parameters are then updated using an optimization algorithm that minimizes the loss. 
This process iterates until all samples in the training set have been used (signaling the end of an epoch), at which point the loss can be reported for the validation set, and the training loop repeated.
The optimization details are very similar to the training routine described in Appendix~A of \cite{WuBoada19}.

We implement and optimize our deep convolutional neural network using \texttt{fastai} \citep{fastai}, which is built on \texttt{PyTorch}. 
All choices of CNN hyperparameters or tweaks have been empirically tuned in order to optimize training.
Performance is primarily quantified by the root mean squared error (RMSE) metric, which also serves as our loss function:

\begin{equation}
    {\rm RMSE} \equiv \sqrt{\langle | \mathcal M_{\rm pred} - \mathcal M_{\rm true}|^2 \rangle }.
\end{equation}
Another important metric of performance is the linear regression slope between $\mathcal M_{\rm true}$ and $\mathcal M_{\rm pred}$.
A slope of unity indicates that there is no regression bias, and a shallower slope generally signifies that the CNN suffers from loss of predictive power (regression attenuation).\footnote{A slope of zero and an RMSE equal to the inherent scatter can be achieved by always predicting the validation sample's mean.}
Other works characterize the scatter using the standard deviation of the difference between predictions and truths, which is systematically lower than the RMSE if there is non-zero mean error (or offset).

We use the xresnet family of CNN architectures, which are enhanced versions of the original residual neural networks \citep{Resnets,BagOfTricks}.
Our 34-layer xresnets are further modified such that the usual Rectified Linear Unit (ReLU) activation functions are replaced with Mish \citep[][]{Mish}, and simple self-attention layers are added after convolutions in the residual blocks \citep{SimpleSelfAttention}. 
We train each model from scratch, as no pretrained CNN with this architecture is available.
In order to iteratively update the CNN's weights, we use a combined Rectified Adam \citep{RAdam} and LookAhead \citep{LookAhead} optimizer. 
Weight decay with a coefficient of $0.01$ is applied to all trainable layers except batch normalization layers \citep{BatchNormWeightDecay}; note that we use true weight decay rather than the L2 norm (see \citealt{WDnotL2} for details).

It is typical to evaluate deep learning models using a validation set drawn from the same distribution as the training set.
If the model performs well on the training data but fails to perform well on the validation data, then it is a sign that the model suffers from overfitting.
We randomly split the data by 80\%/20\% for training/validation sets, unless otherwise noted.

We train batches of 64 images at a time using a Nvidia P100 graphics processing unit (GPU).
The learning rate is scheduled according to the ``one-cycle'' policy for 40 epochs \citep[using the default hyperparameters set by \texttt{fastai};][]{OneCycle}, we set a maximum learning rate of 0.01.
Dihedral group operations are randomly applied to images in order to augment the training set by a factor of eight and to force the CNN to learn such symmetries.
The same transformations are applied during test-time augmentation to the validation data.
We always report RMSE performance for the validation or test set.

We also consider simpler models for regressing $\mathcal M$ based on the image data.
These other regression models are tested in the Appendix~\ref{sec:simple-models}.
Traditional statistical or classical machine learning methods are unable to represent morphological features using pixels as features, while CNNs are designed to encode image shapes, patterns, and textures over multiple scales in a translation- and rotation-invariant way.
In Appendix~\ref{sec:perturbations}, we demonstrate that CNN predictions do not change significantly if the input images are injected with artificial point sources, which reveals that the CNN has learned a robust representation of galaxy morphology.
Thus, we continue our analysis and discussion using the CNN results.

\begin{figure}[t]
    \centering
    \includegraphics[width=\columnwidth]{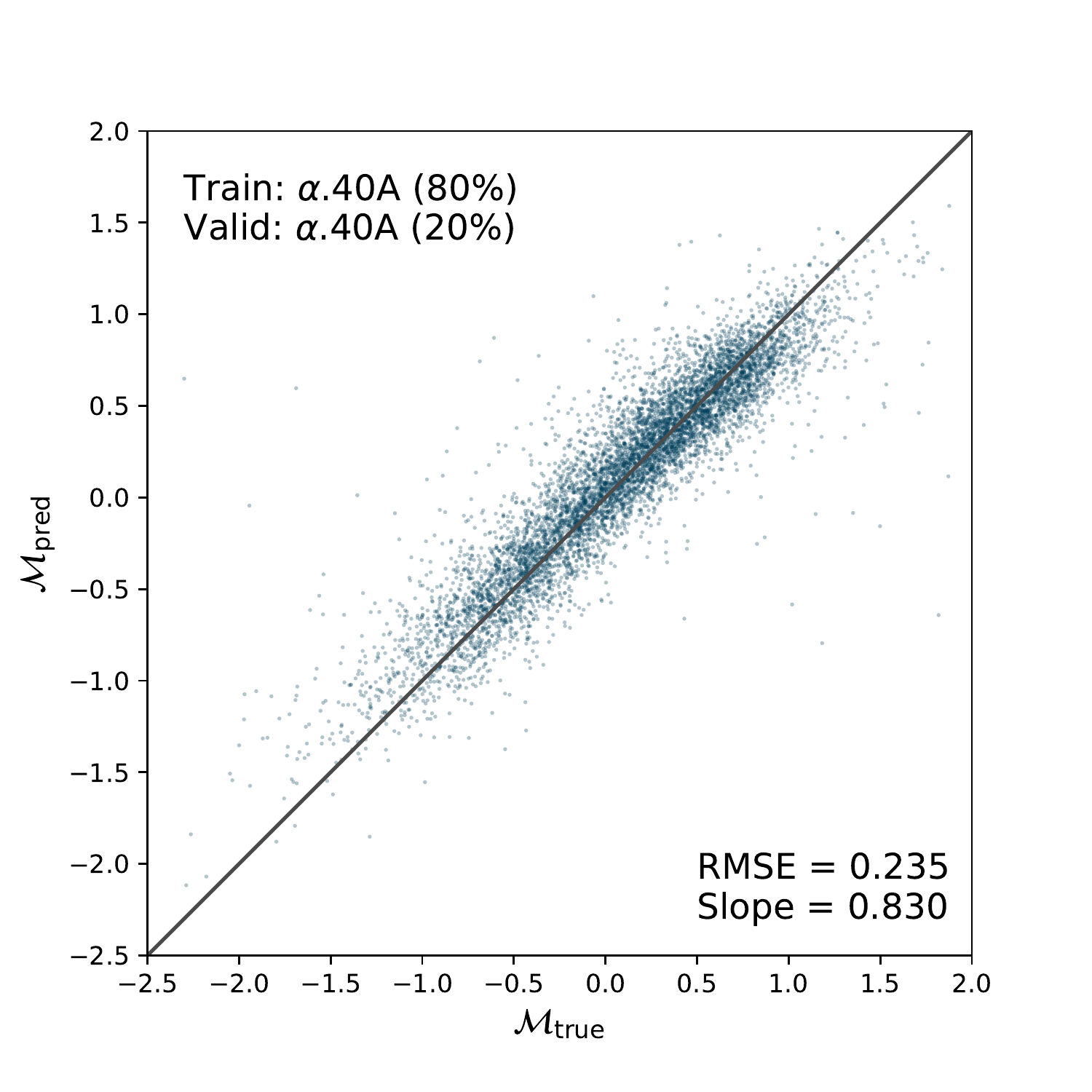}
    \caption{\label{fig:a40A-selfval}
    CNN-predicted logarithmic \HI{} mass fraction ($\mathcal M_{\rm pred}$) plotted against measured values ($\mathcal M_{\rm true}$) for the \alf{}A validation subsample.
    The RMSE loss (in units of dex) and the linear regression slope are shown.
}
\end{figure}

\section{Results} \label{sec:results}

\subsection{Training on \alf{}} \label{sec:self-validation-results}

We train a deep CNN using 80\% of the \alf{}A data set, and evaluate its performance using the remaining 20\% validation data set. 
The optimized model can predict $\mathcal{M}$ to within RMSE$~=0.235~$dex.
In Figure~\ref{fig:a40A-selfval}, we show that the predicted and true values of \HI{} mass fraction agree over two orders of magnitude in $\mathcal M$, and that the slope is close to unity.
Our results demonstrate that there exists a strong relationship between the \HI{} content and the morphological information learned by a CNN from SDSS $gri$ image cutouts.

We repeat the exercise using the smaller \alf{}B sample, and find very similar results (RMSE~$=0.235~$dex), even though \alf{}A is larger than \alf{}B by 54\%.
When we examine the standard deviation of $\mathcal M$ for both subsamples, we find that \alf{}A has much larger scatter ($0.68$~dex) than \alf{}B ($0.50$~dex).
The broader selection criteria for subsample A allows galaxies with poorer constraints on SFR or metallicity, which also means that galaxies with uncommon morphological or \HI{} properties are included.
Therefore, it is not surprising that the smaller \alf{}B subsample, which contains fewer outliers, produces similar results to the larger \alf{}A subsample.\footnote{Training a CNN on the \alf{}A sample leads to better validation performance than training on the full $\alpha$.100 sample (RMSE~$=0.25$~dex). This is likely because the \alf{}A catalog \citep{Haynes+11} is more cleanly matched to SDSS than our custom crossmatching of $\alpha$.100 to SDSS sources \citep{Haynes+18}.}

\subsection{Dependence on galaxy properties}

We find that a trained CNN can accurately recover $\mathcal M$ from optical imaging for the \alf{} data set.
No systematic biases are present, although incorrect predictions tend to be scattered toward the center of the $\mathcal M$ distribution rather than toward the extrema.
Generally, the CNN tends to overpredict gas mass fractions for low-$\mathcal M$ galaxies.

\begin{figure}
    \centering
    \includegraphics[width=1\columnwidth]{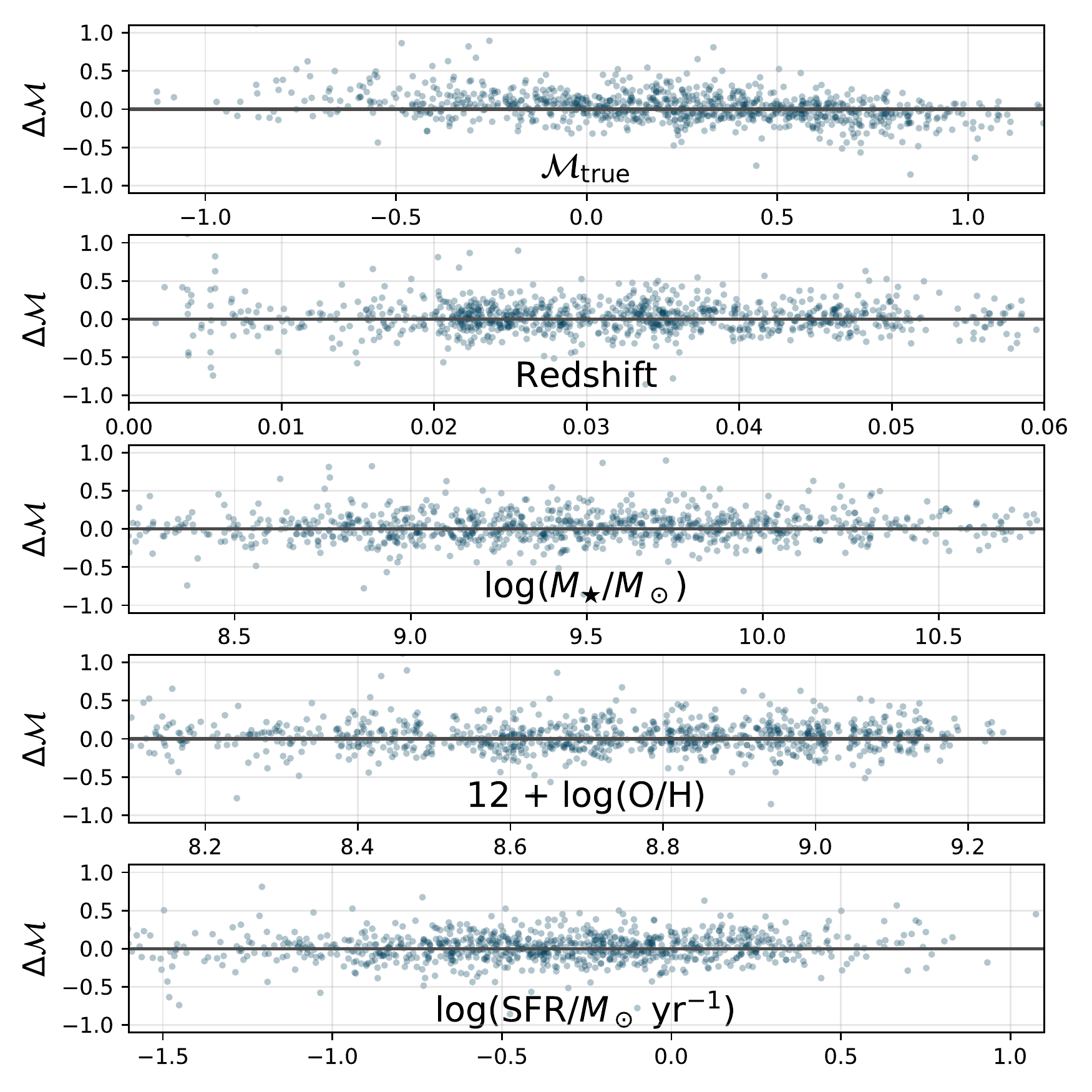}
    \caption{\label{fig:other-properties}
        Comparisons between logarithmic \HI{} mass fraction residuals ($\Delta \mathcal M$) and observed \HI{} mass fraction, redshift, stellar mass, SFR, and gas metallicity.
        Only the \alf{}B validation data are shown.
        Aside from an anti-correlation in $\Delta \mathcal M$ vs $M_{\rm true}$, which appears because the linear regression slope is shallower than unity, no correlation between residuals and other galaxy properties is observed.
    }
\end{figure}

We examine trends between $\Delta \mathcal M \equiv \mathcal M_{\rm pred} - \mathcal M_{\rm true}$ and other physical properties of galaxies.
For example, it may be that the CNN tends to under- or overpredict $\mathcal M$ based on some image feature that also correlates with other galaxy properties. 
However, we find that $\Delta \mathcal M$ does not vary systematically with any other property, and nor does the amount of scatter in $\Delta \mathcal M$.
The only trend that we find is a negative correlation between $\Delta \mathcal M$ and $\mathcal M_{\rm true}$, which is expected when the regression slope is less than unity.
In Figure~\ref{fig:other-properties}, we show trends between $\Delta \mathcal M$ and $\mathcal M_{\rm true}$, redshift, stellar mass, SFR, and gas metallicity for the \alf{}B validation data set (959 \HI{} sources).
$\Delta \mathcal M$ also does not correlate with specific SFR, 4000~\AA{} break strength, or the $\delta_5$ environmental parameter (discussed in Section~\ref{sec:environment}).

\subsection{Testing on $\alpha$.100, NIBLES, and xGASS} \label{sec:test-results}

\begin{figure}
    \centering
    \includegraphics[height=0.8\textheight]{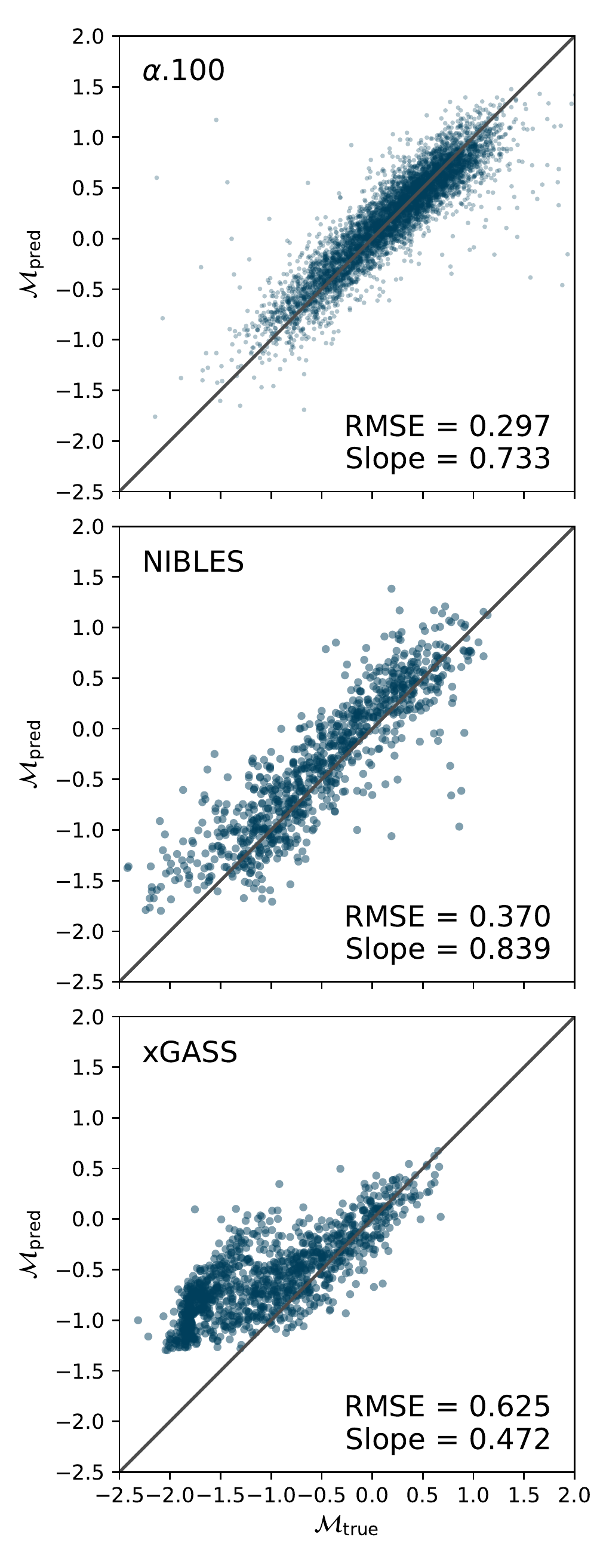}
    \caption{\label{fig:initial-CNN-tests}
    CNN-predicted logarithmic \HI{} mass fraction ($\mathcal M_{\rm pred}$) plotted against measured values ($\mathcal M_{\rm true}$) for the $\alpha$.100, NIBLES, and xGASS test samples.
}
\end{figure}

We use the CNN trained on \alf{}A to estimate $\mathcal M_{\rm pred}$ for galaxies in the $\alpha$.100, NIBLES, and xGASS test data sets.
In the first set of entries in Table~\ref{tab:CNN-PR-results}, we report our results for each data set.
We list the RMSE, slope, $\sigma$, and mean offset in order to quantify performance, although we primarily rely on the first two metrics.
CNN predictions are also shown in Figure~\ref{fig:initial-CNN-tests}.

\subsubsection{$\alpha$.100 results} \label{sec:a100-results}

By construction, our $\alpha$.100 test sample does not overlap with \alf{}, so it can be used as an independent test for the trained CNN.
In the top panel of Figure~\ref{fig:initial-CNN-tests}, we show a scatter plot of $\mathcal M_{\rm pred}$ against $\mathcal M_{\rm true}$ for $\alpha$.100.
Overall, we find that the CNN recovers $\mathcal M_{\rm true}$ accurately to within 0.30~dex.
The performance is weaker than the \alf{} validation sample (RMSE~$=0.235$~dex), which implies that the CNN is unable to perfectly generalize to the $\alpha$.100 catalog.
However, we note that our custom crossmatch of $\alpha$.100 to SDSS counterparts is different from the published \alf{} crossmatch to SDSS counterparts \citep{Haynes+11}, and that disparate selection effects may cause issues in generalization as well.
One consequence of the imperfect crossmatching is that several sources apparently have implausibly high gas mass fractions.
When 43 sources with $\mathcal M_{\rm true} > 2 $ are removed from the $\alpha$.100 sample, we find that the test performance is RMSE~$=0.23~$dex, which is in agreement with our \alf{} training sample.
We will discuss the impacts of additional selection effects in Section~\ref{sec:T17-comparison}.

\subsubsection{NIBLES results}

The NIBLES catalog presents another opportunity to test our CNN trained on \alf{}A. 
While the two samples have comparable stellar mass distributions, NIBLES extends to significantly lower gas mass fractions than \alf{}.
We find that the CNN is generally able to recover $\mathcal M_{\rm true}$ to within $0.37~$dex (see center panel of Figure~\ref{fig:initial-CNN-tests}).
We note that $\mathcal M_{\rm pred}$ estimates for NIBLES are systematically high by about 0.14~dex (Table~\ref{tab:CNN-PR-results}).
Our results are consistent with previous findings that ALFALFA \HI{} fluxes are about $0.16$~dex (45\%) higher than NIBLES measurements for the overlapping sample \citep[possibly due to a combination of flux calibration differences and multi-beam flux reconstruction;][]{NIBLES}.
If we rescale \HI{} masses by the $0.16$~dex systematic offset, then we find that the CNN's predictions are in better agreement with NIBLES measurements (RMSE~$=0.34~$dex).

\subsubsection{xGASS results}

The xGASS representative galaxy sample contains a considerable number of massive elliptical galaxies with little gas content (see Table~\ref{tab:summary}).
For these gas-poor systems, the CNN tends to overpredict $\mathcal M$ because it has been trained on \alf{}A, which comprises mostly \HI{}-rich, star-forming galaxies.
We observe that the CNN does not output values below $\mathcal M_{\rm pred} \approx -1.3$ for xGASS (although this does not appear to be a severe problem for NIBLES).
As a result, our metrics indicate that the RMSE is large and the slope is too flat, as shown in the bottom panel of Figure~\ref{fig:initial-CNN-tests}.
Therefore, the CNN is unable to generalize to \textit{out-of-distribution} galaxies such as massive, gas-poor systems in xGASS.

\section{Pattern recognition for out-of-distribution samples} \label{sec:PR}

\subsection{Out-of-distribution samples} \label{sec:out-of-distribution}

The trained CNN is capable of making accurate predictions where the training and test set distributions of $\mathcal M_{\rm true}$ overlap.
For example, over 95\% of the \alf{}A training sample have $\mathcal M_{\rm true} > -1$, and it is evident from Figure~\ref{fig:initial-CNN-tests} that the CNN's predictions for xGASS are more accurate for higher values of $\mathcal M_{\rm true}$.
However, for galaxies without measured \HI{} masses, it is not known \textit{a priori} whether a galaxy's $\mathcal M_{\rm true}$ is within the distribution of the training sample, and we must consider whether a galaxy is out-of-distribution in the input space (image data) rather than the target space ($\mathcal M_{\rm true}$).
A CNN (or any regression algorithm) is essentially a mapping between the input and output space, so it is feasible that the out-of-distribution $\mathcal M_{\rm true}$ can be related to out-of-distribution galaxy images.
For the rest of the paper, we will consider \alf{}A detections to be ``in-distribution'' and ALFALFA non-detections to be ``out-of-distribution'' for the trained CNN.

A neural network can be trained to recognize patterns in order to classify galaxies as in- or out-of-distribution \citep{Hopfield87,Bishop95,Kinney+96}.
Pattern recognition (PR) algorithms have been used extensively in astronomy \citep[e.g.,][]{Zhu+14,Teimoorinia+17,Caldeira+19,Ciprijanovic+20}, and they can supplement other machine learning estimators by classifying whether an input is representative of a training distribution.
One such example is presented by \cite{Teimoorinia+17}, where a shallow neural network is trained to distinguish ALFALFA \HI{} detections from non-detections.
Their method relies on 15 galaxy parameters derived from SDSS spectroscopy and photometry.
However, one of the restrictions with this approach is that it necessitates spectroscopic observations and ancillary data in order to make predictions. 
Our method only requires an image of the galaxy.
We proceed by training a CNN PR algorithm based on optical imaging in order to distinguish ALFALFA detections from non-detections.

\subsection{Pattern recognition with a CNN}

We obtain $gri$ imaging for a sample of $z < 0.06$ SDSS galaxies that are located in the ALFALFA footprint but are undetected in the $\alpha$.100 catalog.
To ensure a clean sample of ALFALFA non-detections, we impose the same selection cuts used for $\alpha$.100 (Section~\ref{sec:a100}), and remove all galaxies with neighboring \HI{} sources within an Arecibo beam radius.
We randomly select 7,399 ALFALFA non-detections from this parent sample to enforce balanced classes with \alf{}A detections.

We train a CNN PR algorithm using exactly the same architecture as before, except that the final layer now predicts two outputs.
After applying a sigmoid function, these two outputs represent the probabilities of detection ($p_{\rm CNN}$) and non-detection ($1 - p_{\rm CNN}$) in an ALFALFA-like survey.
The \alf{}A sample serves as ground truths for the ALFALFA detection category, and the non-detection sample serves as ground truths for the non-detection category.
We use the same optimization routine as before, except that cross-entropy loss is used as the optimization function since the objective now is binary classification. 
We also apply label smoothing with $\epsilon = 0.05$, so that optimized values of $p_{\rm CNN}$ gravitate toward 0.05 and 0.95 for non-detections and detections, respectively (see, e.g., \citealt{LabelSmoothing}).
We optimize the CNN using the same hyperparameters discussed in Section~\ref{sec:CNNs}.

\subsection{PR results}

\begin{figure}
    \centering
    \includegraphics[width=\columnwidth]{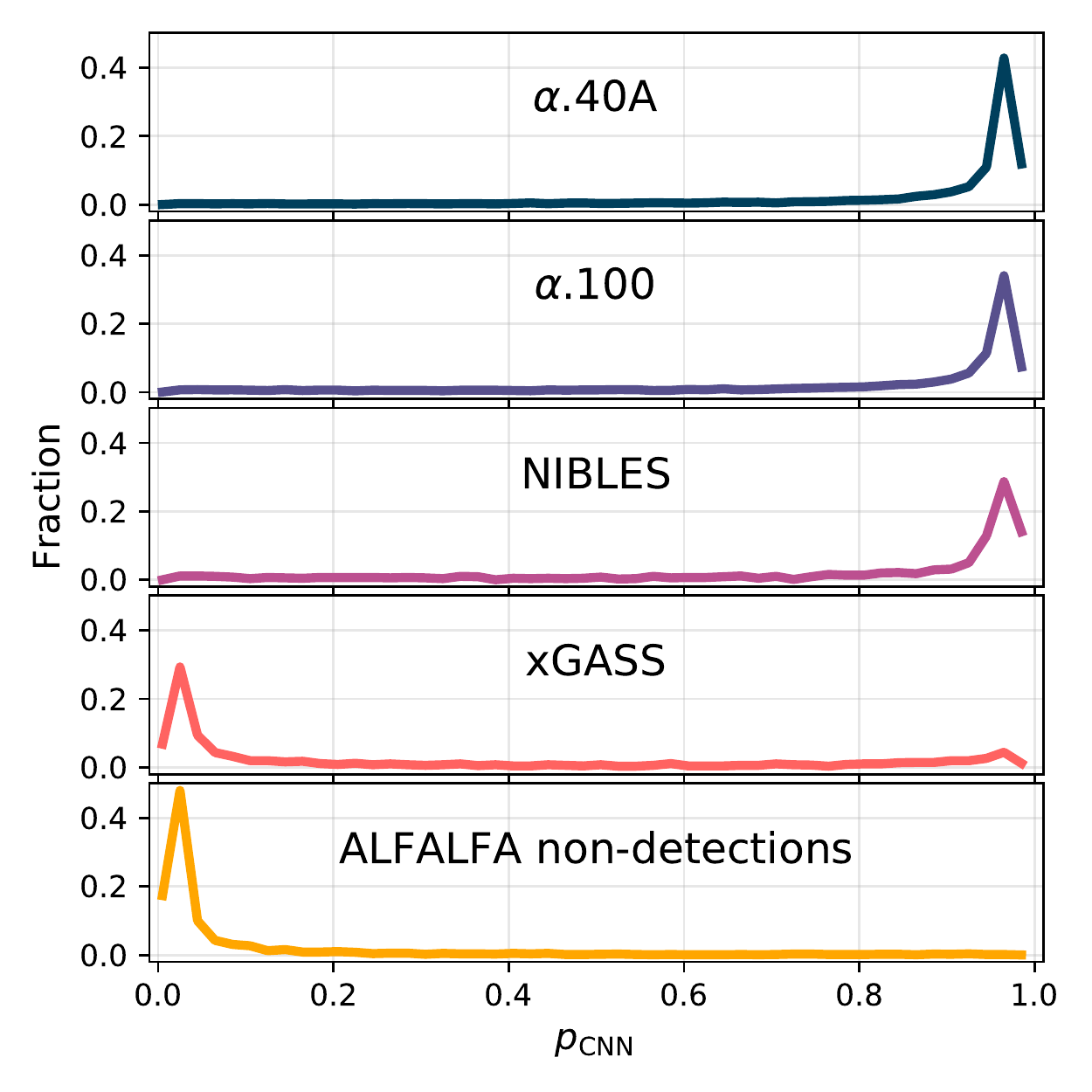}
    \caption{
    Pattern recognition probability ($p_{\rm CNN}$) distributions for different galaxy samples.
    The \alf{}A and ALFALFA non-detections samples are used to train the CNN, while the $\alpha$.100, NIBLES, and xGASS are test galaxy samples.
    }
    \label{fig:PR-results}
\end{figure}

In the top and bottom panels of Figure~\ref{fig:PR-results}, we show the distributions of $p_{\rm CNN}$ for the \alf{}A and ALFALFA non-detection samples, i.e., the training subsamples. 
Their $p_{\rm CNN}$ distributions are strongly peaked at $\sim 0.95$ and $0.05$, respectively, indicating that the CNN robustly identifies \alf{}A detections solely using optical imaging.
The trained PR is able to distinguish ALFALFA detections from non-detections with 93\% accuracy and AUC~$=0.93$ (area under the curve for the receiver operating characteristic).\footnote{
The receiver operating characteristic (ROC) curve evaluates a model's performance at all classification decision boundaries. Generally, the false positive rate is plotted against the true positive rate, such that the expected area under the curve (AUC) is 0.5 for a completely random model with balanced classes, and the AUC~$=1$ for a perfectly accurate and precise model.
}

We examine the $p_{\rm CNN}$ distributions for the $\alpha$.100, NIBLES, and xGASS samples, which are also shown in Figure~\ref{fig:PR-results}.
As expected, the $\alpha$.100 galaxies are predominantly characterized by high values of $p_{\rm CNN}$.
Most galaxies in the NIBLES sample are also at $p_{\rm CNN} \approx 0.95$ and would be detected by an ALFALFA-like survey.
The xGASS representative galaxy sample, conversely, is characterized mostly by low values of $p_{\rm CNN}$, although there is a small fraction with high $p_{\rm CNN}$.
Our results verify that the gas mass fraction estimates suffered from out-of-distribution error when evaluated on the xGASS sample, but did not encounter the same issue for the $\alpha$.100 and NIBLES samples.

\subsection{Combining $\mathcal M_{\rm pred}$ and $p_{\rm CNN}$} \label{sec:detailed-results}

We can use the CNN PR results to assess the reliability of $\mathcal M_{\rm pred}$ on the test data.
We consider various threshold values of $p_{\rm CNN}$ to separate detections from non-detections.
$p_{\rm CNN} > 0.5$ represents a natural decision boundary, although in practice we may find that other values are better.
For example, we have trained the CNN PR using an equal ratio of detected and undetected galaxies, but the ALFALFA survey only detects about $20\%$ of typical galaxies in the low-redshift Universe \citep{Catinella+10}, which may imply that a higher value of $p_{\rm CNN}$ is desirable for rejecting false detections in typical massive galaxy samples.

In Table~\ref{tab:CNN-PR-results}, we present CNN regression results for all \HI{} data sets using several choices of PR decision boundary.
For the ALFALFA data sets, we find that different cuts in $p_{\rm CNN}$ does not significantly impact regression performance (as characterized by, e.g., RMSE).
This is expected because the ALFALFA samples should not contain out-of-distribution examples, and thus the PR cuts will not strongly affect the results.\footnote{There is a subtle effect with the $\alpha.100$ test set that causes the RMSE to slightly \textit{increase} as we restrict $p_{\rm CNN}$ to higher values. Galaxies labeled with lower $p_{\rm CNN}$ generally have moderate \HI{} properties (i.e., they are near the mode of the $\mathcal M_{\rm true}$ distribution), whereas sources labeled with higher values of $p_{\rm CNN}$ might be extremely \HI{}-rich (i.e., they have a broader distribution of $\mathcal M_{\rm true}$). The end result is a $< 0.01$~dex increase as we shift the decision boundary from $p_{\rm CNN}>0.5$ to $0.9$.}
For NIBLES, performance modestly improves as the $p_{\rm CNN}$ threshold increases from zero (effectively the same as no cut) to $0.5$, $0.8$, and $0.9$.
We find that strict $p_{\rm CNN}$ cuts remove gas-poor (e.g., $\mathcal M_{\rm true} < -1$) NIBLES systems for which $\mathcal M_{\rm pred}$ is systematically overestimated.
The most dramatic improvement is seen in xGASS, for which a PR cut of $p_{\rm CNN} > 0.5$ removes nearly three quarters of the sample.
Increasing the $p_{\rm CNN}$ threshold further refines the CNN performance; the RMSE, $\sigma$, and mean offset decrease, and the slope increases.
For $p_{\rm CNN} > 0.9$, the xGASS test set RMSE ($0.24$~dex) is comparable to that of the \alf{}A training set ($0.23$~dex).

\begin{deluxetable}{c l r RRRR}[ht]
    \tablewidth{0pt}
    \tablecolumns{7}
    \tabletypesize{\scriptsize}
    \tablecaption{Combined pattern recognition and $\mathcal M_{\rm pred}$ results \label{tab:CNN-PR-results}}
    \tablehead{
        \colhead{PR cut} &
        \colhead{Data set} & 
        \colhead{$N$} &
        \colhead{RMSE} & 
        \colhead{Slope} &
        \colhead{$\sigma$} & 
        \colhead{Offset} \vspace{-0.5em} \\
        \colhead{} & 
        \colhead{} & 
        \colhead{} & 
        \colhead{(dex)} & 
        \colhead{} & 
        \colhead{(dex)} & 
        \colhead{}
    }
    \startdata
    \hline
    & $\alpha$.40A & 7128 & 0.2335 & 0.8427 &  0.2318 & 0.0285  \\
    & $\alpha$.100 & 6087 & 0.2975 & 0.7325 & 0.2974 & -0.0076  \vspace{-0.75em}\\
    \vspace{-0.75em} None & \\
    & NIBLES & 899 & 0.3705 & 0.8395 & 0.3416 & 0.1438 \\
    & xGASS & 1179 & 0.6254 & 0.4724 & 0.4284 & 0.4558 \\
    \hline
    &$\alpha$.40A & 6674 & 0.2320 &  0.8429 & 0.2302 & 0.0291 \\
    & $\alpha$.100 & 5249 & 0.3043 & 0.7192 & 0.3042 & -0.0086 \vspace{-0.75em} \\
    \vspace{-0.75em} $p_{\rm CNN} > 0.5$ & \\
    & NIBLES & 767 & 0.3566 & 0.8518 & 0.3238 & 0.1499 \\
    & xGASS  & 326 & 0.3237 & 0.6423 & 0.2962 & 0.1314 \\
    \hline
    & $\alpha$.40A & 6004 & 0.2334 & 0.8434 & 0.2316 & 0.0293 \\
    & $\alpha$.100 & 4454 & 0.3123 & 0.7084 & 0.3122 & -0.0077 \vspace{-0.75em} \\
    \vspace{-0.75em} $p_{\rm CNN} > 0.8$ & \\
    & NIBLES & 663 & 0.3508 & 0.8586 & 0.3170 &  0.1506 \\
    & xGASS  & 217 & 0.2755 & 0.6882 & 0.2669 & 0.0706 \\
    \hline
    & $\alpha$.40A &  5319 & 0.2322 & 0.8468 & 0.2304 & 0.0290 \\
    & $\alpha$.100 &  3787 & 0.3107 & 0.7116 & 0.3107 & -0.0085 \vspace{-0.75em} \\
    \vspace{-0.75em} $p_{\rm CNN} > 0.9$ & \\
    & NIBLES & 572 & 0.3435 & 0.8703 & 0.3116 & 0.1451 \\
    & xGASS  & 143 & 0.2449 & 0.7486 & 0.2365 & 0.0667  
    \enddata
    \tablecomments{
    The scatter can be quantified using the RMSE or standard deviation ($\sigma$) metrics, where lower is better. 
    We also list the regression slope (closer to unity is better) and average offset (closer to zero is better).
    All data sets are independent from each other, such that $N$ is sometimes smaller than the sample sizes listed in Table~\ref{tab:summary}.
    }
\end{deluxetable}

A larger threshold for $p_{\rm CNN}$ enforces higher purity of ALFALFA-like galaxies.
We recommend that $p_{\rm CNN}>0.9$ be used for typical massive galaxy samples similar to the xGASS representative sample.
However, purity comes at the cost of completeness, and we note that different science goals may call for a different balance between complete versus clean samples.
For optically selected samples, we find that a $p_{\rm CNN} > 0.9$ threshold can robustly remove \HI{} non-detections.

\section{A comparison of $\mathcal M$ estimators} \label{sec:T17-comparison}

\begin{figure*}
    \centering
    \includegraphics[height=0.8\textheight]{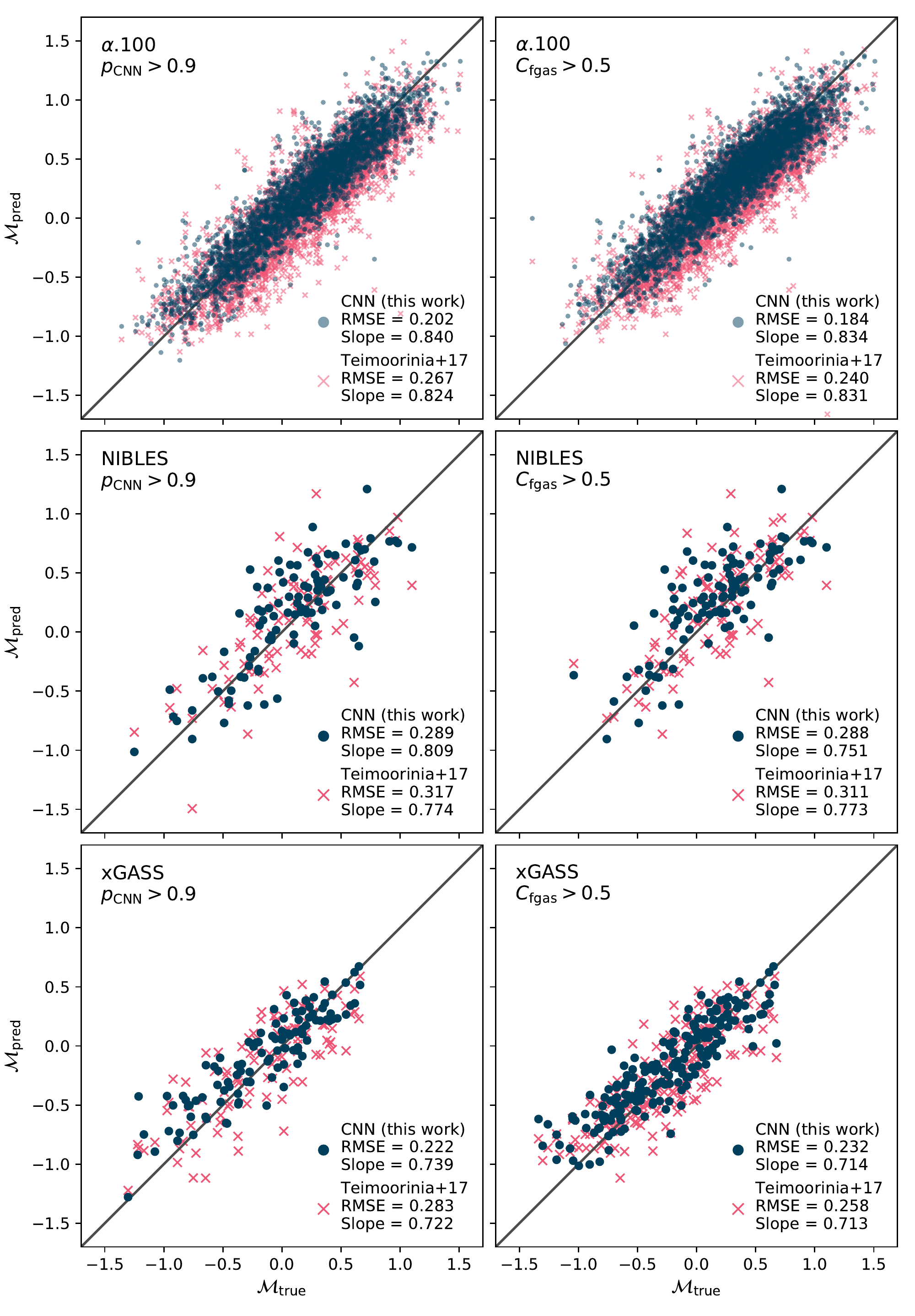}
    \caption{
    Comparison of $\mathcal M_{\rm pred}$ versus $\mathcal M_{\rm true}$ for our CNN trained on \alf{}A (blue circles) and the fully connected neural network presented \cite[pink crosses;][]{Teimoorinia+17}.
    We show results for the $\alpha$.100 (top), NIBLES (middle), and xGASS (bottom) samples, using a selection of $p_{\rm CNN} > 0.9$ (left) or $C_{\rm fgas} > 0.5$ (right).
    Performance metrics are listed in each panel.
    }
    \label{fig:scatter-plot-T17-comparisons}
\end{figure*}

\subsection{Colors and morphological parameters}

Many works have studied the correlations between galaxy properties and their \HI{} content.
\cite{Kannappan04} finds that optical and near-infrared colors are connected to gas-to-stellar mass fraction, and reports a relationship between $u-K$ and $\mathcal M$ with only $\sigma = 0.37$~dex scatter.
\cite{Zhang+09} use $i$-band surface brightness, $\mu_i$, in addition to $g-r$ color to predict $\mathcal M$.
They find that the best-fit relation reduces the scatter to $0.31$~dex.
Other morphological parameters have been shown to tighten the relationship between galaxy colors and $\mathcal M$, such as stellar mass surface density and axial ratios, which reduce $\sigma \sim 0.30~$dex \citep[e.g.,][]{Catinella+10,Huang+12,Li+12,Catinella+13,Eckert+15}.

\cite{Teimoorinia+17} extend the \cite{Zhang+09} linear method by fitting a quadratic estimator to the inputs.
They report a best fit of $\sigma = 0.22$~dex on the training data; however, this estimator is unable to generalize well to gas-poor samples such as GASS.
Conversely, \cite{Catinella+10} predict $\mathcal M$ for the GASS sample using UV-optical colors and stellar mass surface density, but their method begins to systematically fail in the gas-rich regime of very blue galaxies.
The lack of generalizability suggests that more robust estimators are needed, and that pattern recognition should be used to exclude galaxies that yield poor predictions \citep[e.g.,][]{Catinella+13}.

\subsection{Machine learning methods}

Machine learning has recently become more popular for constructing powerful and flexible $\mathcal M$ estimators.
\cite{Rafieferantsoa+18} estimate gas mass fraction for observed and simulated galaxy samples by using a variety of machine learning algorithms, such as random forests, gradient-boosted trees, and deep neural networks. 
Their algorithms can achieve RMSE~$= 0.25-0.3$~dex when trained trained on real data (using photometric and environmental parameters as input features), but the estimators are unable to accurately predict $\mathcal M$ when trained on simulated data \citep[see also][]{Andrianomena+20}.
We also train classical machine learning algorithms to model three-color \textit{image} inputs, and demonstrate that they predict $\mathcal M$ to RMSE~$= 0.31$~dex for \alf{}A (Appendix~\ref{sec:simple-models}).

\cite{Teimoorinia+17} show that a fully connected neural network (FCNN) can be used to estimate $\mathcal M$ to $\sigma = 0.22$~dex on independent test sets when trained using 15 galaxy properties.
Although several of these input features are known to individually covary with $\mathcal M$, the non-linear combination of these properties processed by a shallow neural network is able to robustly outperform traditional regression methods (see \citealt{Ellison+16} for details on the network architecture).
They find that $g-r$ color and $\mu_i$ are the most important parameters for regression (similar to previous results), and that a galaxy's bulge-to-total fraction also plays a role in governing $\mathcal M$.
\cite{Teimoorinia+17} also present a pattern recognition method for rejecting galaxies that are not representative of the training sample and tend to be incorrectly predicted.
The authors combine the PR probability and the epistemic uncertainty determined from an ensemble of neural network predictions, $\sigma_{\rm fitN}$, to form $C_{\rm fgas}$, which parameterizes the robustness of their $\mathcal M$ prediction.
\cite{Teimoorinia+17} have publically released $\mathcal M$ and $C_{\rm fgas}$ estimates for a sample of over 500,000 SDSS galaxies.

\subsection{Comparison to \cite{Teimoorinia+17}}

The FCNN by \cite{Teimoorinia+17} predicts $\mathcal M$ with remarkably low scatter.
Because their method outperforms previous photometric gas fraction techniques \citep[e.g.,][]{Kannappan04,Zhang+09}, their results serve as a valuable baseline for comparison to our work.
In order to make a fair comparison with their results, we first crossmatch our data to their catalogs using a 1\arcsec{} radius.
This process removes a non-trivial fraction of the test set before any PR cuts are made; after crossmatching, 67\% of \alf{}A, 87\% of $\alpha$.100, 18\% of NIBLES, and 86\% of xGASS remain.

\begin{figure*}
    \centering
    \includegraphics[width=\textwidth]{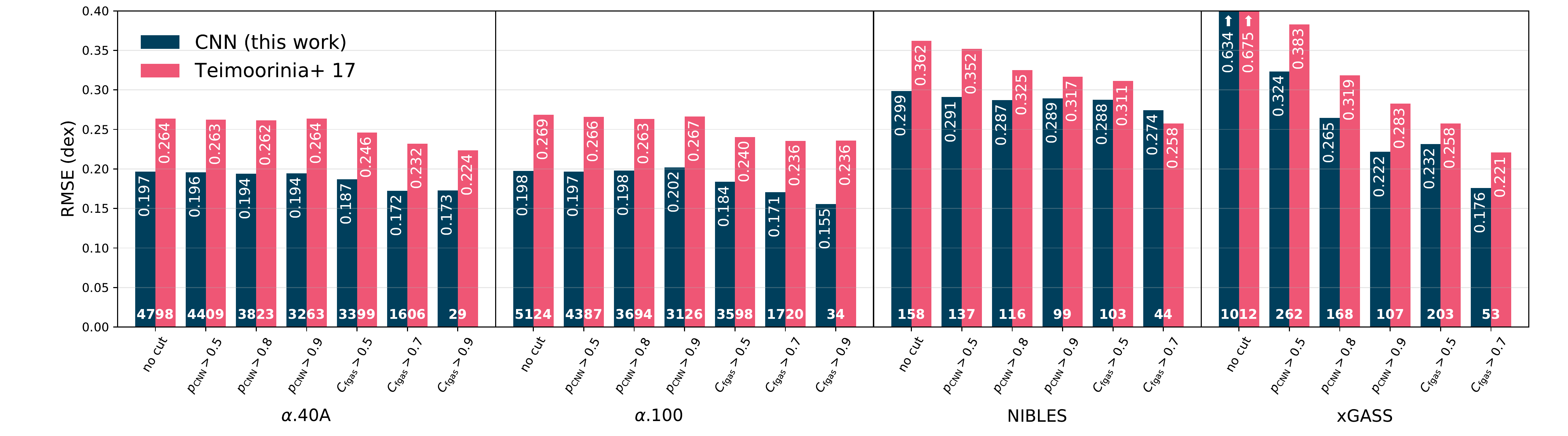}
    \caption{
    Comparison of results using our CNN method (\textit{blue}) and a fully connected neural network \citep[\textit{red};][]{Teimoorinia+17} for ALFALFA, xGASS, and NIBLES galaxies crossmatched with the \cite{Teimoorinia+17} catalog.
    We show the RMSE for test subsamples selected using various choices of pattern recognition ($C_{\rm fgas}$ and $p_{\rm CNN}$) or no cut at all; in every case the CNN recovers $\mathcal M$ with lower scatter.
    The number of galaxies in each test set is shown in at the bottom of each bar plot.
    Detailed comparisons with additional metrics are provided in Table~\ref{tab:CNN-T17-comparison}.
}
    \label{fig:T17-comparison}
\end{figure*}

In Figure~\ref{fig:T17-comparison}, we show $\mathcal M_{\rm pred}$ versus $\mathcal M_{\rm true}$ scatter plots comparing our predictions with \cite{Teimoorinia+17} results. 
Overall, we find that the CNN outperforms the FCNN according in terms of scatter and slope.
Using a conservative threshold of $p_{\rm CNN} > 0.9$, we observe that the CNN predicts $\mathcal M$ to RMSE~$=0.20$~dex for $\alpha$.100, compared to 0.27~dex for the FCNN.
If we instead apply $C_{\rm fgas} > 0.5$ to both the CNN and \cite{Teimoorinia+17} results, we find RMSE~$=0.18$ and $0.24$~dex, respectively, for $\alpha$.100.
Both methods give comparable results for the crossmatched NIBLES catalog (RMSE~$=0.29$~dex for the CNN, and $0.31$~dex for the FCNN).
For xGASS, we find that the CNN predictions are robust using either $p_{\rm CNN}>0.9$ (RMSE~$=0.22$~dex) or $C_{\rm fgas} > 0.5$ (RMSE~$=0.23$~dex), whereas the FCNN performs better using the latter ($0.26$~dex).

\begin{deluxetable*}{clr c RRRR c RRRR}
    \tablewidth{0pt}
    \tablecolumns{13}
    \tablecaption{Comparison of CNN and FCNN Results \label{tab:CNN-T17-comparison}}
    \tablehead{
        \colhead{PR cut} &
        \colhead{Data set} & 
        \colhead{$N$} &&
        \multicolumn{4}{c}{CNN (this work)} &&
        \multicolumn{4}{c}{FCNN \citep{Teimoorinia+17}} \\
        \cline{5-8} \cline{10-13}
        &&&&
        \colhead{RMSE (dex)} & 
        \colhead{Slope} &
        \colhead{$\sigma$ (dex)} & 
        \colhead{Offset} &&
        \colhead{RMSE (dex)} & 
        \colhead{Slope} &
        \colhead{$\sigma$ (dex)} & 
        \colhead{Offset} 
    }
    \startdata
    \hline
    & $\alpha$.40A & 4798 &&  \textbf{0.1967} & \textbf{0.8524} & \textbf{0.1953} & \textbf{0.0238} && 0.2637 & 0.8317 & 0.2556 & -0.0647 \\
    & $\alpha$.100 & 5124 && \textbf{0.1975} & \textbf{0.8462} & \textbf{0.1971} & \textbf{0.0125} && 0.2688 & 0.8243 & 0.2574 & -0.0775    \vspace{-0.63em}\\
    \vspace{-0.63em} None & \\
    & NIBLES & 158 &  &  \textbf{0.2988} & \textbf{0.7890} &  \textbf{0.2791} & 0.1090 && 0.3622 & 0.7194 & 0.3474 & \textbf{0.1060} \\
    & xGASS & 1012 &  & \textbf{0.6338} & \textbf{0.4584} & \textbf{0.4258} & \textbf{0.4697} &  & 0.6751 & 0.4098 & 0.4793 & 0.4758 \\
    \hline
    &$\alpha$.40A & 4409 &  & \textbf{0.1958} & \textbf{0.8531} & \textbf{0.1942} & \textbf{0.0247} &  & 0.2626 & 0.8297 & 0.2548 & -0.0636 \\
    & $\alpha$.100 & 4387 &  & \textbf{0.1968} & \textbf{0.8473} & \textbf{0.1963} & \textbf{0.0138} &  & 0.2660 &  0.8267 & 0.2550 &  -0.0758 \vspace{-0.63em} \\
    \vspace{-0.63em} $p_{\rm CNN} > 0.5$ & \\
    & NIBLES & 137 &  &  \textbf{0.2909} & \textbf{0.7898} & \textbf{0.2739} & 0.1009 &  &  0.3518 & 0.7208 & 0.3435 & \textbf{0.0815} \\
    & xGASS  & 262 &  &  \textbf{0.3235} & \textbf{0.6221} & \textbf{0.2953} & 0.1333 &  &  0.3829 & 0.5796 & 0.3649 & \textbf{0.1182} \\
    \hline
    & $\alpha$.40A & 3823 &  & \textbf{0.1938} & \textbf{0.8574} & \textbf{0.1922} & \textbf{0.0250} &  & 0.2617 & 0.8343 & 0.2530 &  -0.0672 \\
    & $\alpha$.100 & 3694 &  & \textbf{0.1980} &  \textbf{0.8440} &  \textbf{0.1974} & \textbf{0.0160}  &  & 0.2634 & 0.8263 & 0.2521 & -0.0765 \vspace{-0.63em} \\
    \vspace{-0.63em} $p_{\rm CNN} > 0.8$ & \\
    & NIBLES & 116 &  &  \textbf{0.2869} & \textbf{0.7981} & \textbf{0.2760} &  0.0826 &  &  0.3252 & 0.7361 & 0.3226 & \textbf{0.0509} \\
    & xGASS  & 168 &  &  \textbf{0.2645} & \textbf{0.6602} & \textbf{0.2568} & 0.0665  &  &  0.3185 & 0.6566 & 0.3182 & \textbf{0.0283} \\
    \hline
    & $\alpha$.40A &  3263 &  & \textbf{0.1944} & \textbf{0.8589} & \textbf{0.1927} & \textbf{0.0261} &  & 0.2638 & 0.8340 &  0.2552 & -0.0670 \\
    & $\alpha$.100 & 3126 &  & \textbf{0.2021} & \textbf{0.8405} & \textbf{0.2015} & \textbf{0.0160} &  & 0.2665 & 0.8239 & 0.2548 & -0.0783 \vspace{-0.63em} \\
    \vspace{-0.63em} $p_{\rm CNN} > 0.9$ & \\
    & NIBLES & 99 &  & \textbf{0.2895} & \textbf{0.8090} &  \textbf{0.2822} & 0.0706  &  & 0.3167 & 0.7737 & 0.3174 & \textbf{0.0234} \\
    & xGASS  & 107 &  &  \textbf{0.2216} & \textbf{0.7392} & \textbf{0.2098} & 0.0742 &  &  0.2826 & 0.7222 & 0.2835 & \textbf{0.0145} \\
    \hline
    & $\alpha$.40A &  3399 &  & \textbf{0.1869} & \textbf{0.8306} & \textbf{0.1863} & \textbf{0.0157}  &  & 0.2460 &  0.8179 & 0.2347 & -0.0738 \\
    & $\alpha$.100 & 3598 &  & \textbf{0.1838} & \textbf{0.8344} & \textbf{0.1837} & \textbf{0.0063} &  & 0.2404 & 0.8311 & 0.2243 & -0.0867 \vspace{-0.63em} \\
    \vspace{-0.63em} $C_{\rm fgas} > 0.5$ & \\
    & NIBLES & 103 &  &  \textbf{0.2876} & 0.7507 & \textbf{0.2692} & 0.1044 &  &  0.3113 & \textbf{0.7734} & 0.3091 & \textbf{0.0476} \\
    & xGASS  & 203 &  &  \textbf{0.2317} & \textbf{0.7138} & \textbf{0.2189} & 0.0777 &  &  0.2577 & 0.7127 & 0.2581 & \textbf{0.0112} \\
    \hline 
    & $\alpha$.40A & 1606 &  & \textbf{0.1725} & \textbf{0.8142} & \textbf{0.1725} & \textbf{0.0039}  &  & 0.2321 &  0.7968 & 0.2083 & -0.1025 \\
    & $\alpha$.100 & 1720 &  & \textbf{0.1706} & 0.8238 & \textbf{0.1707} & \textbf{-0.0009} &  & 0.2356 &  \textbf{0.8318} & 0.2074 & -0.1119 \vspace{-0.63em} \\
    \vspace{-0.63em} $C_{\rm fgas} > 0.7$ & \\
    & NIBLES & 44 &  &  0.2744 & 0.5946 & \textbf{0.2532} & 0.1124 &  &  \textbf{0.2577} & \textbf{0.7070} & 0.2605 & \textbf{-0.0109} \\
    & xGASS  & 53 &  &  \textbf{0.1758} & \textbf{0.7999} & \textbf{0.1589} & 0.1124 &  &  0.2208 & 0.7494 & 0.2134 & \textbf{-0.0640} \\
    \hline 
    & $\alpha$.40A & 29 &  & \textbf{0.1730} & \textbf{0.6614} & 0.1735 & \textbf{-0.0290}  &  & 0.2236 &  0.6599 & \textbf{0.1666} & -0.1523 
    \vspace{-0.63em} \\
    \vspace{-0.63em} $C_{\rm fgas} > 0.9$ & \\ 
    & $\alpha$.100 & 34 &  & \textbf{0.1555} & \textbf{0.7549} & 0.1524 & \textbf{-0.0405} &  & 0.2358 &  0.7454 & \textbf{0.1446} & -0.1879 
    \enddata
    \tablecomments{
    We compare $\mathcal M_{\rm pred}$ from this work and from the fully connected neural network trained by \cite{Teimoorinia+17} using all data sets and several choices of pattern recognition cuts.
    In order to facilitate an equal comparison, only galaxies with matches in the \cite{Teimoorinia+17} catalog are evaluated.
    Bolded values indicate superior performance for a given combination of PR cut, data set, and metric.
    }
\end{deluxetable*}

In Table~\ref{tab:CNN-T17-comparison}, we provide detailed comparisons of our results with those published by \cite{Teimoorinia+17}.
We test different $p_{\rm CNN}$ and $C_{\rm fgas}$ thresholds in addition to the $C_{\rm fgas} > 0.5$ threshold recommended by \cite{Teimoorinia+17}.
For all data sets, the CNN performs extremely well under the $p_{\rm CNN}$ and $C_{\rm fgas}$ selection criteria.
However, it is difficult to compare the CNN and FCNN results for $C_{\rm fgas}$ thresholds higher than 0.5. 
For example, a $C_{\rm fgas} > 0.9$ cut removes all NIBLES and xGASS galaxies and only leaves a few galaxies in alpha.40 ($N=29$) and alpha.100 ($34$).
Even with a more moderate cut ($C_{\rm fgas} > 0.7$), the sample size is small for NIBLES ($44$) and xGASS ($53$).
Therefore, our conclusions are based on the $C_{\rm fgas} = 0.5$ decision boundary.

We primarily quantify our results using the RMSE scatter.
However, previous works often report $\sigma$ when discussing scatter \citep{Zhang+09,Teimoorinia+17}. 
When there is no mean offset between $\mathcal M_{\rm pred}$ and $\mathcal M_{\rm true}$, $\sigma$ is comparable to the RMSE. 
However, the RMSE will more heavily penalize predictions when there is an offset, such as the 0.14~dex offset in \HI{} mass between NIBLES and ALFALFA, or the $>0.20$~dex systematic overpredictions for \HI{}-poor galaxies in xGASS (prior to applying PR cuts).
We recommend using $\sigma$ for comparison only when the mean offset is small compared to the scatter.\footnote{It is worth noting that \cite{Teimoorinia+17} systematically predict lower $\mathcal M$ than we do, which causes negative offsets in their ALFALFA predictions relative to $\mathcal M_{\rm true}$.
Equivalently, our NIBLES and xGASS predictions have positive offsets relative to $\mathcal M_{\rm true}$.
The systematic offset between our CNN and their FCNN predictions is likely due to their strict exclusion of galaxies with neighboring \HI{} sources, which leads to a 0.14~dex difference in $\mathcal M_{\rm true}$ between their clean and contaminated samples.}
Nevertheless, we find that the CNN consistently outperforms the FCNN according to both RMSE and $\sigma$.

It is intriguing that the CNN results are significantly improved after crossmatching with the \cite{Teimoorinia+17} catalog (i.e., comparing RMSE in Tables~\ref{tab:CNN-PR-results} and \ref{tab:CNN-T17-comparison}).
Although we previously find RMSE$~=0.30$~dex for the $\alpha$.100 test set, we now report RMSE$~=0.20$~dex, in part because the sources with unrealistic $\mathcal M_{\rm true}$ values identified in Section~\ref{sec:a100-results} have have been removed.
Substantial improvements are also evident for the other data sets.
This discrepancy is most likely due to selection criteria introduced by \cite{Teimoorinia+17}, which can remove galaxies with uncertain physical properties.
A similar selection effect accounts for why \alf{}B has smaller intrinsic scatter in $\mathcal M$ than \alf{}A.
\cite{Teimoorinia+17} restrict their analysis to galaxies with magnitudes in all SDSS bands, redshifts, sizes, morphological measurements, and environmental parameters.
Galaxies with all 15 properties are characterized by higher signal-to-noise observations and are more likely to have secure \HI{} and stellar mass measurements.
The omitted galaxies tend to be redder and therefore more likely to contribute error and scatter.
We conclude that the combination of selection criteria introduced during the comparison with \cite{Teimoorinia+17} explains the major improvement in our CNN predictions.

\section{The impact of environment} \label{sec:environment}

The morphology and \HI{} properties of a galaxy are strongly sensitive to its surrounding environment \citep[e.g.,][]{Serra+12,Jones+16}.
If a CNN can accurately learn a connection between galaxy optical imaging and $\mathcal M$ when trained only on systems in clustered environments but fails to accurately estimate \HI{} content for a test sample of isolated galaxies (or vice versa), then it may be a sign that the distribution of galaxy morphologies has shifted. 
In essence, we wish to probe whether the \HI{}-morphology relation learned by the CNN covaries with environment.

\subsection{Galaxy overdensity}\label{sec:overdensity}

\begin{figure}[t]
    \centering
    \includegraphics[width=\columnwidth]{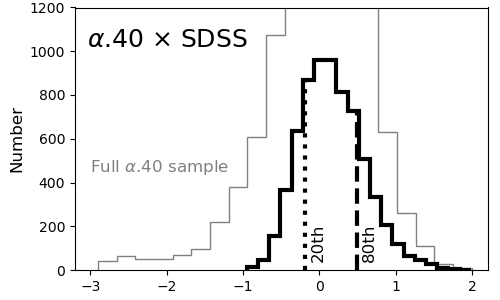}
    \caption{\label{fig:environment}
        Thick black lines show histogram distributions of normalized galaxy overdensity for the \alf{} sample.
        Dotted and dashed vertical lines in both panels represent the 20th and 80th percentile values for $\delta_5$, respectively. 
        We also show the parent \alf{} sample in light gray (prior to optical crossmatching with the SDSS catalog).
        }
\end{figure}

\begin{figure*}[t]
    \centering
    \includegraphics[width=\textwidth]{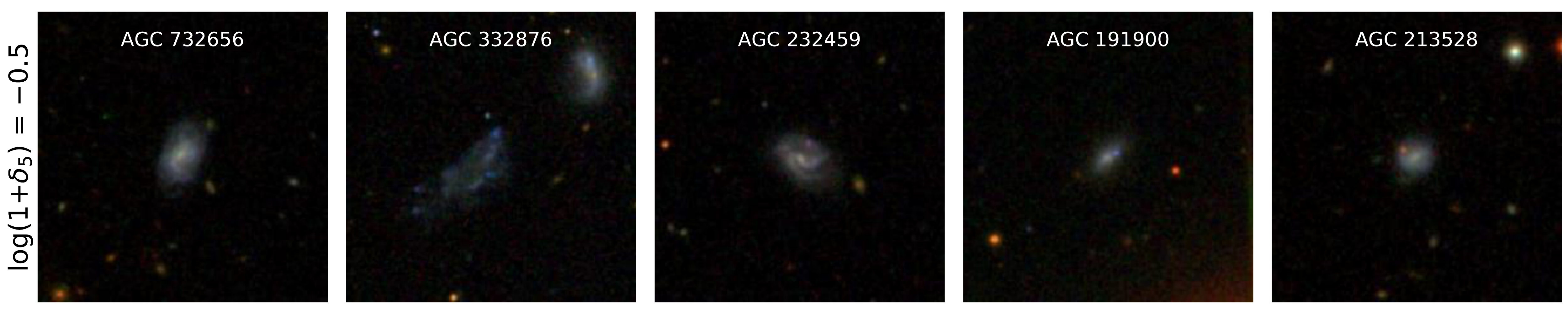}
    \includegraphics[width=\textwidth]{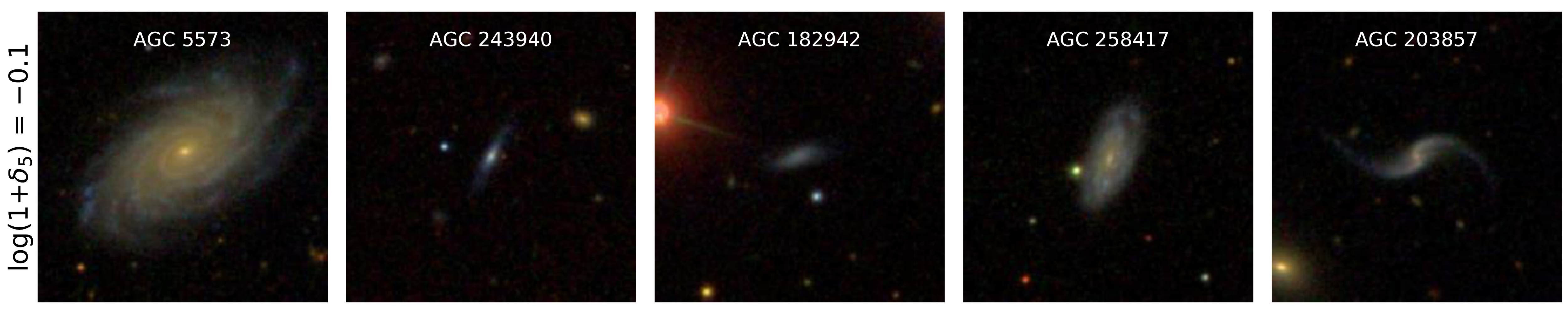}
    \includegraphics[width=\textwidth]{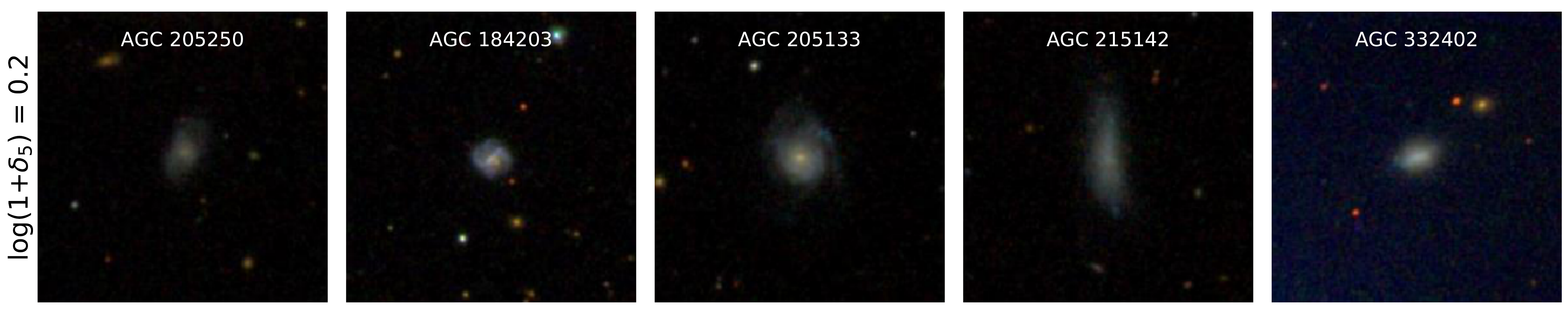}
    \includegraphics[width=\textwidth]{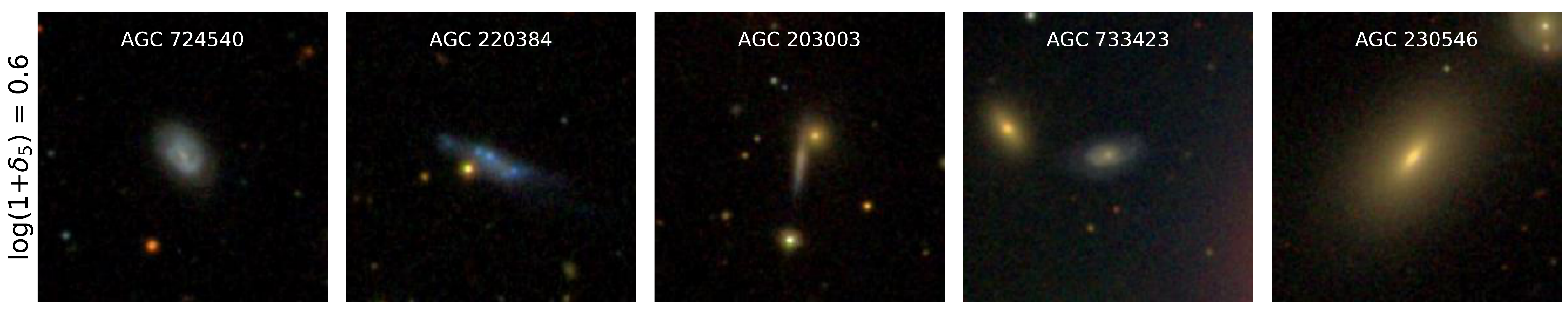}
    \includegraphics[width=\textwidth]{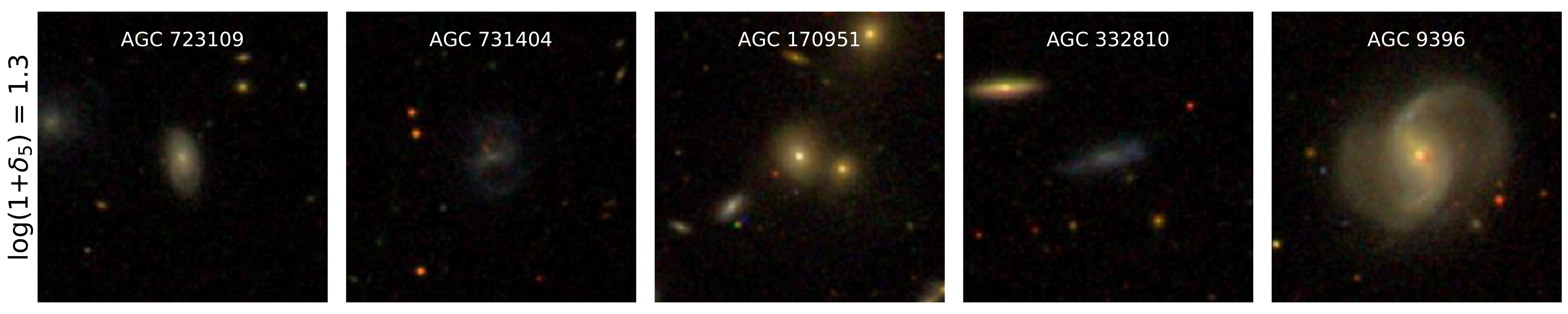}
    \caption{
    Example SDSS images of \alf{}A galaxy image cutouts in different environmental regimes.
    Each row depicts five galaxies with environmental overdensity closest to the value indicated on the left. 
    The example galaxies range from below 10th percentile to over 90th percentile in $\log (1 + \delta_5)$.
    }
    \label{fig:delta_5-examples}
\end{figure*}

In order to quantitatively investigate the impacts of environment, we parameterize the environment using the projected galaxy density \citep[e.g.,][]{Cooper+08}:
\begin{equation}
    \Sigma_5 = \frac{3}{\pi D_5^2},
\end{equation}
where $D_5$ is the projected physical distance to each galaxy's fifth-nearest neighbor (including its own optical counterpart).
We use neighboring galaxies in the NASA-Sloan Atlas (version 1.0.1; \citealt{NSAtlas}) within a velocity window of $\pm \rm 1000~km~s^{-1}$ in order to compute $\Sigma_5$ for each \HI{} source.
We enforce a $D > 10$~Mpc distance cut in order to prevent contamination or biases from the Local Group.
Following \cite{Cooper+08}, we divide $\Sigma_5$ by its median over a sliding redshift boxcar window of size $\Delta z = 0.02$, which removes redshift dependence.
The final result is a normalized overdensity parameter, $1+\delta_5$.
In Figure~\ref{fig:environment}, we show the distribution of $\log(1+\delta_5)$ for our \alf{} sample crossmatched with spectroscopically confirmed SDSS optical counterparts (we also show the full \alf{} sample in gray).
It is apparent that the optical-\HI{} crossmatching exercise removes \alf{} systems in the lowest-density environments.
In Figure~\ref{fig:delta_5-examples}, we provide examples of SDSS image cutouts for ALFALFA galaxies at various environmental densities.

We select 80\% of the higher-$\delta_5$ galaxies for training and set aside the remaining 20\% (with lower $\delta_5$) for validation.
In other words, we test whether a CNN trained on galaxies in higher-density environments can accurately predict the \HI{} content of galaxies in lower-density environments.
If this turns out to be the case, all else unchanged, then the environment does not significantly impact the connection between \HI{} richness and optical imaging learned by the CNN.
We also split the sample such that the 80\% with lower $\delta_5$ is used for training, and the 20\% with higher $\delta_5$ is used for validation.
As a baseline comparison, we test the case where the training and validation set are randomly split (but trained in the same manner otherwise).\footnote{For each environmental test run, a 34-layer xresnet is trained for 10 epochs using a learning rate of 0.03, batch size of 32, weight decay of $10^{-4}$, and the validation subsample is evaluated using test-time augmentation. These hyperparameters have been chosen to best optimize the CNN in a smaller number of training epochs so that we can run multiple tests quickly.}
We repeat tests five times for each training/validation split, and report the RMSE average and standard deviation in Table~\ref{tab:environment}.

\begin{deluxetable*}{ll ccc}
    \tablewidth{0pt}
    \tablecolumns{5}
    \tablecaption{\alf{}A results split by environment \label{tab:environment}}
    \tablehead{
        \colhead{Training} & 
        \colhead{Validation} & 
        \colhead{Validation $\sigma$} & \colhead{Validation RMSE} & 
        \colhead{Normalized RMSE} \vspace{-0.5em}\\
        \colhead{$N=5922$} &
        \colhead{$N=1477$} &
        \colhead{(dex)} &
        \colhead{(dex)} &
        \colhead{}
    }
    \startdata
        $[0.2, 1.0)$ & $[0, 0.2)$ &  $0.5241$ & $0.2184 \pm 0.0022$ & $0.4167 \pm 0.0042$ \\
        $[0, 0.8)$ & $[0.8, 1.0)$ & $0.6706$ & $0.3269 \pm 0.0066$ & $0.4874 \pm 0.0098$ \\
        Random & Random  & $0.6036$ & $0.2557 \pm 0.0094$ & $0.4237 \pm 0.0156$
    \enddata
    \tablecomments{
    Comparison of CNN performance using different training/validation splits for \alf{}A.
    The training and validation subsamples are either randomly selected or separated by $\delta_5$ quantile according to an 80\%/20\% ratio in the given quantile range.
    }
\end{deluxetable*}

Our initial tests suggest that the galaxy \HI{}-morphology connection \textit{varies significantly} with environment.
We find that a CNN trained only on galaxies in overdense environments and validated on systems in underdense environments performs better than the inverse.
Surprisingly, the CNN validated on lower-$\delta_5$ systems even outperforms the randomized baseline.
However, this effect is fully explained by the validation scatter in $\mathcal M$ when we select subsamples by a range in $\delta_5$.
When we compare the CNN performance normalized by the inherent scatter of the validation subsample (the right-most column in Table~\ref{tab:environment}), it becomes apparent that the CNN validated on higher-$\delta_5$ environments still performs significantly worse than those validated on random or lower-$\delta_5$ environments.
We conclude that the CNN is able to generalize predictions in a way that yields good performance in underdense environments when exposed to galaxies in more overdense environments (relative to randomized validation subsamples), yet the opposite is not true.

These results can be interpreted as evidence that the \HI{}-morphology connection is controlled by different physical mechanisms in the highest-$\delta_5$ environments.
Galaxies residing in clusters are subject to ram pressure stripping, tidal forces, galaxy-galaxy interactions, and other effects that can leave morphological imprints and also depress their gas content \citep[e.g.,][]{Chung+09,Fabello+12}. 
By training on subsamples of galaxies characterized by relatively lower-density surroundings, a CNN is unlikely to learn about the morphological cues associated with extreme physics of clustered environments, and therefore our experiment is able to distinguish between ``typical'' and ``overdense'' modes of the \HI{}-morphology connection.
It is also worth noting that these tests may not even capture the full extent of the environmental effects, as the 3.8\arcmin{} Arecibo beam may cause overestimation of \HI{} mass or misidentification of optical counterparts in groups and clusters \citep[e.g.,][]{Serra+15,Stevens+19}.
Such errors may artificially boost \HI{} content and thereby ameliorate the CNN's performance in high-density environments.

\subsection{The overdensity transition regime} \label{sec:environmental-transition}

\begin{figure}[t]
    \centering
    \includegraphics[width=\columnwidth]{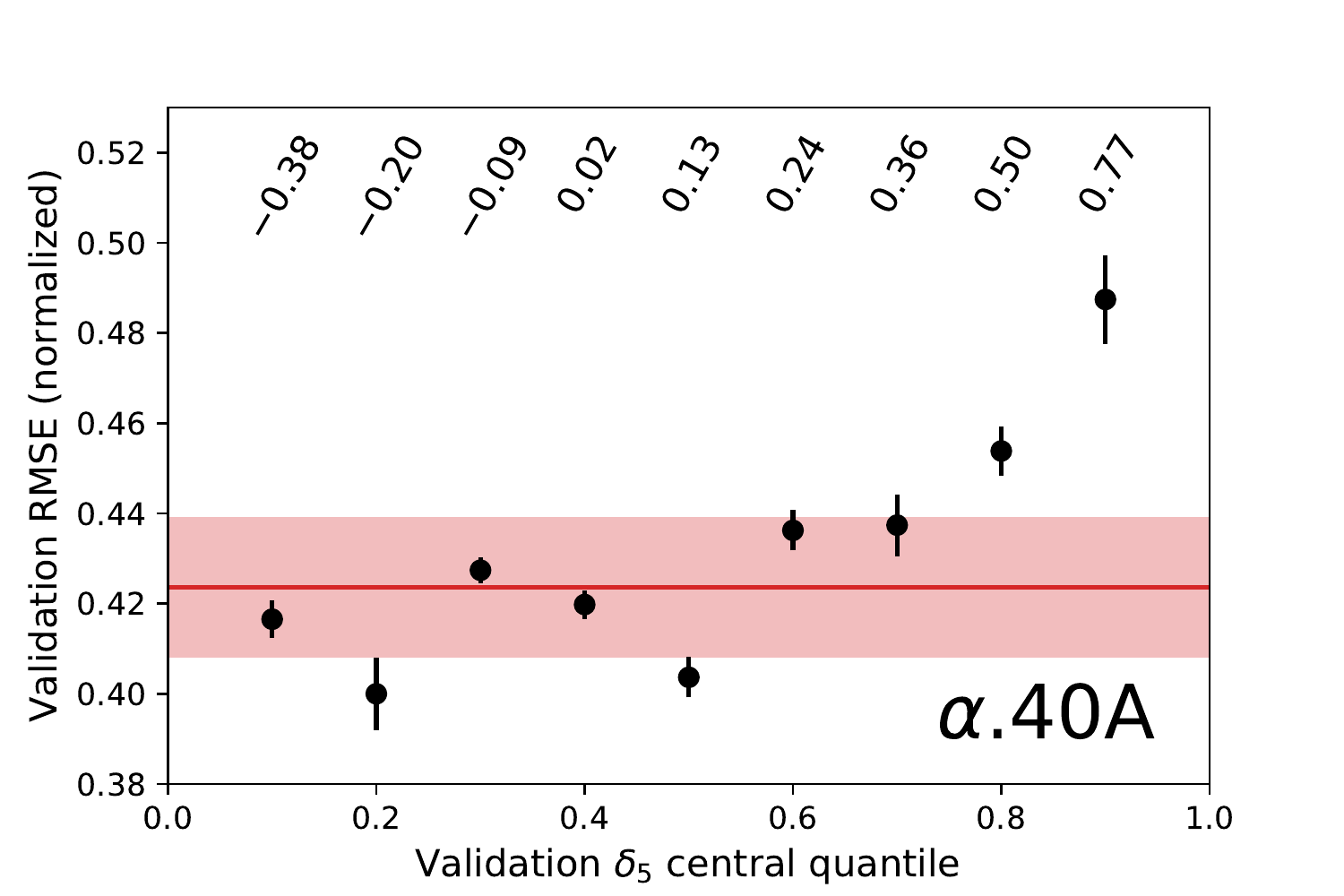}
    \caption{
        CNN validation performance across different environmental densities shown in black markers and error bars.
        The performance is the RMSE normalized by the inherent scatter in $\mathcal M$ for the validation set; we show the mean and standard deviation for five tests.
        Each validation set is constructed from a 20\% range in $\delta_5$, and the corresponding central $\log(1+\delta_5)$ value is shown at the top.
        Validation results from randomly drawn subsamples are shown in red.
    }
    \label{fig:environment-transition}
\end{figure}

We observe a stark difference in CNN performance across different density regimes, but it is unlikely that there is a sharp transition in environmental effects.
Gas mass fraction is known to depend on a satellite galaxy's distance toward the center of its group or cluster host \citep[e.g.,][]{Brown+17}.
``Pre-processing'' in only moderately overdense environments can also depress galaxies' gas masses \citep{Odekon+16}.
We probe the gradual onset of environmental effects by repeating the analysis in Section~\ref{sec:overdensity} and reserving 20\% of the galaxies for validation based on their $\delta_5$.
We show the normalized RMSE as a function of the validation set $\delta_5$ in Figure~\ref{fig:environment-transition}.
For example, one of the validation data sets in \alf{}A comprises galaxies with $\delta_5$ values in the $0.1 - 0.3$ quantile range, and the training set would consist of the remainder of the sample.
The central validation $\delta_5$ quantile is $0.2$, corresponding to a value of $\log(1+\delta_5) = -0.20$), and the normalized RMSE is approximately $0.40 \pm 0.01$.

We find that the \HI{}-morphology relation transitions to a different ``mode'' at high $\delta_5$.
For low-density environments, a CNN is able to leverage the morphological information learned in intermediate- and high-density regimes and accurately predict the gas mass fraction directly from imaging.
For high-density environments, a CNN is not able to generalize information learned from low- and intermediate-density regimes as well, and the normalized RMSE increases significantly.
A physical explanation for this transition is the growing importance of ram pressure stripping, tidal forces, and other gas depletion effects in overdense environments.
We determine that these effects become increasingly significant at $0.8$ quantile in $\delta_5$ for \alf{}A, corresponding to a normalized overdensity of $\log (1+\delta_5) \geq 0.5$; for lower values of $\delta_5$, the physics that govern this \HI{}-morphology relation are constant.

\section{Interpreting morphological features} \label{sec:interpretation}

Deep learning models often contain millions of trainable parameters, which makes them difficult to interpret compared to classical statistical or smaller machine learning models.
For this reason, deep neural networks are often viewed as opaque systems that cannot be trusted.
Indeed, there are many cases in which deep learning algorithms make high-confidence predictions while failing spectacularly; e.g., when a CNN is confronted with adversarial examples, out-of-distribution predictions, or domain adaptation problems \citep[e.g.,][]{AI-Safety}.
In Section~\ref{sec:test-results}, we examined how a CNN trained on ALFALFA could not be used to make predictions for xGASS without pattern recognition, because the two galaxy populations have different distributions of physical properties.
Alternatively, poor generalization across domains can occur if the training and test data sets are systematically different, e.g., if the training data comprises images of simulated galaxies while the test set solely comprises images of observed galaxies \citep[e.g.,][]{Rafieferantsoa+18,Andrianomena+20}.
In Section~\ref{sec:environment}, we exploited these failure modes in order to probe how galaxy environment impacts the learned \HI{}-morphology relation.
However, other unknown factors may cause CNNs to perform poorly. 
Therefore, it is critical to investigate whether or not trained CNNs make sensible predictions in line with our physical intuition.

It is possible to decipher deep CNNs by examining which parts of an image are most relevant for making certain predictions.
This method of localizing image features is generally known as \textit{input attribution}, because output predictions can be directly attributed to pixels from the input images.
Input attribution methods such as saliency maps and class activation maps \citep{Saliency,CAM} are useful for identifying the image features that a trained CNN ``looks at'' in order to make its predictions.
Other methods can also provide valuable insights, such as by visualizing the learned convolutional layers in optimzied CNNs \citep[e.g.,][]{VisualizingCNNs}.

\begin{figure*}[ht]
    \begin{center}
    \includegraphics[width=0.497\textwidth]{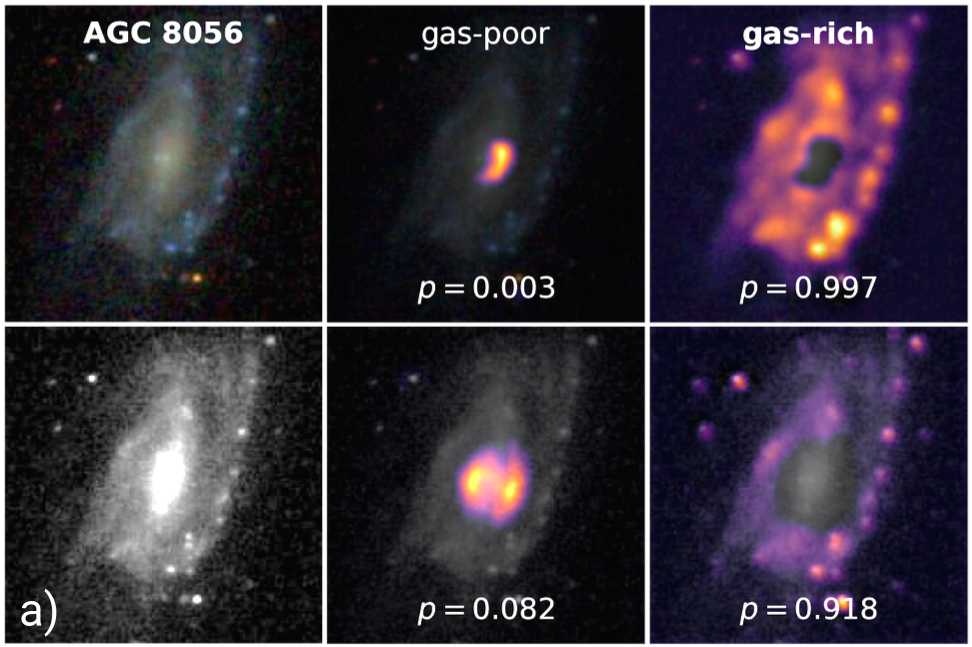}\hspace{-0.2em}
    \includegraphics[width=0.497\textwidth]{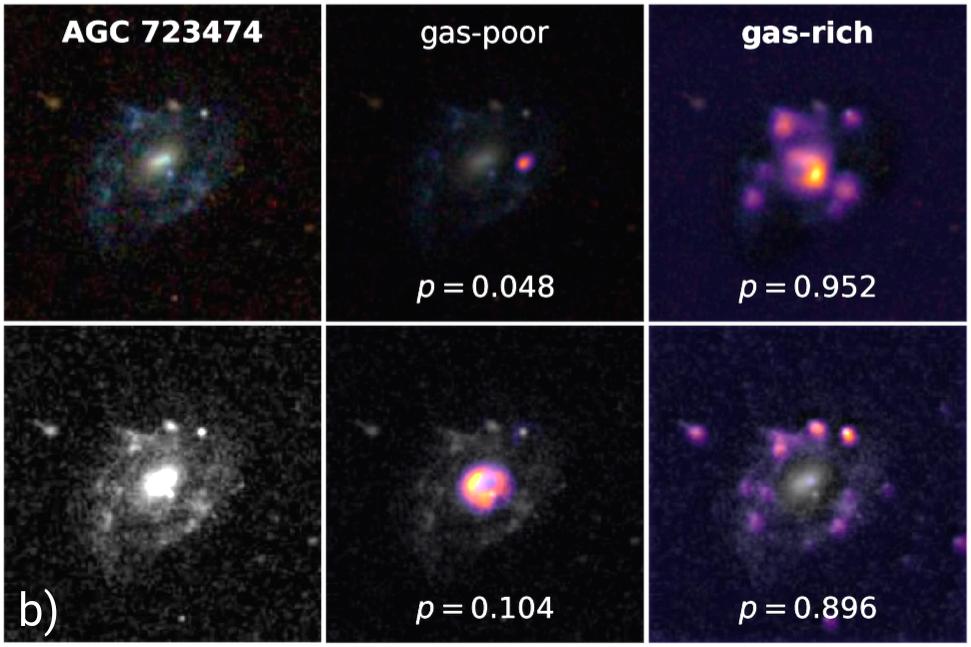} 
    \includegraphics[width=0.497\textwidth]{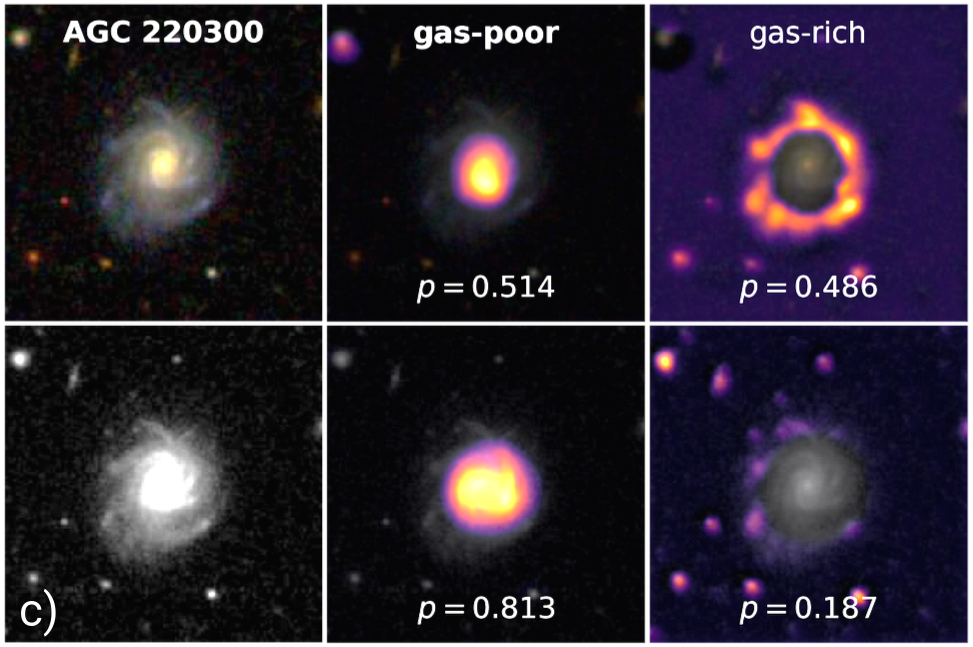}\hspace{-0.2em}
    \includegraphics[width=0.497\textwidth]{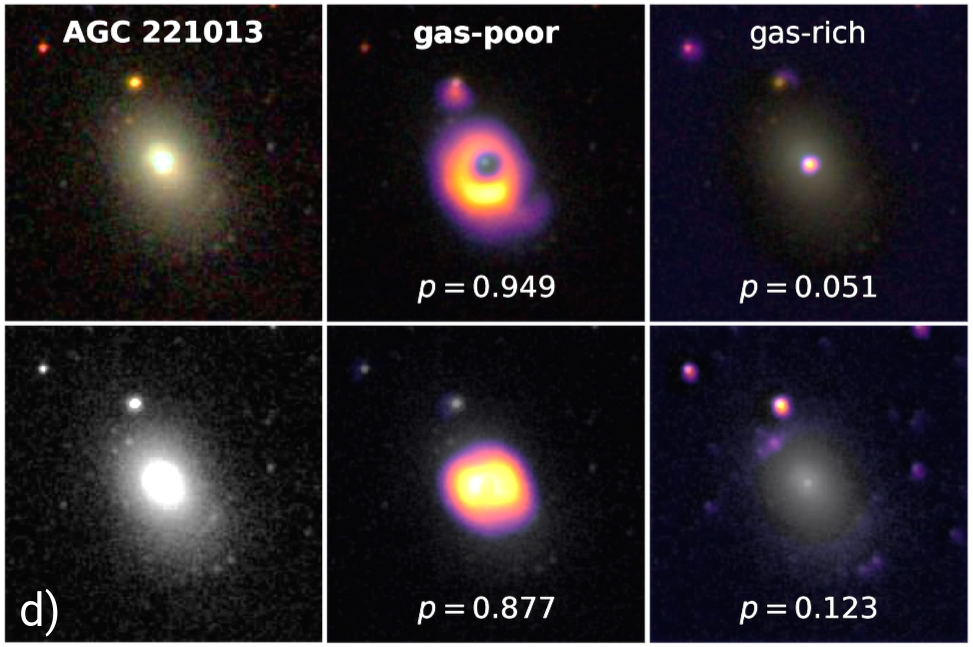} 
    \caption{\label{fig:grad-cam}
        Grad-CAM heatmaps shown for SDSS images using trained CNNs.
        Each panel shows, from left to right, the SDSS $gri$ image cutout, the heatmap of activations corresponding to gas-poor features, and the heatmap of activations corresponding to gas-rich features.
        Grad-CAM heatmaps are shown for for $gri$ color input images (upper) and monochromatic input images (lower).
        Gas-poor/gas-rich labels are bolded for ground truth values, and CNN probabilities are provided for each class. 
        The image contrast has been reduced for visualization purposes.
    }
    \end{center}
\end{figure*}

\subsection{Grad-CAM}
To interpret our results, we make use of the Gradient-weighted Class Activation Map (Grad-CAM) visualization algorithm \citep{GradCAM}. 
Grad-CAM is an input attribution tool that highlights the activated ``neurons'' in a trained CNN corresponding to the pixels in an input image that are most important for predicting a designated class.
These visual explanations enable us to directly attribute output predictions to input morphological features represented as pixels.
In order to use this algorithm, we reformulate our gas mass fraction regression problem to a binary classification problem.

We train a CNN to classify gas-rich and gas-poor galaxies in the \alf{}A sample.
We define low-$\mathcal M$ (gas-poor) and high-$\mathcal M$ (gas-rich) as $\mathcal M < -0.5$ (1,327 objects) and $\mathcal M > 0.5$ (1,922 objects) respectively, so that the two classes are well-separated.
According to this classification, some star-forming galaxies are labeled as ``gas-poor,'' so this scheme is only appropriate for the \HI{}-rich ALFALFA sample.

We also use a simple CNN for visualization purposes.
Our previous architecture (34-layer xresnet) contains many pooling layers that each decrease the resolution by a factor of two, such that the final Grad-CAM result is a $7\times 7$ pixel feature map.
Instead, we use a shallower CNN that consists of a basic CNN stem and two residual blocks containing two convolutional layers each \citep[see, e.g.,][]{fastai}.
The final convolutional layer outputs a $56 \times 56$ pixel feature map.
We use the same optimization methods as in Section~\ref{sec:CNNs}, except that we train for only 10 epochs (at which point we reach convergence) and optimize using cross entropy loss with $\epsilon = 0.05$ label smoothing.
The shallow network classifies \alf{}A galaxies by gas richness with 98\% accuracy.

\subsection{The most important morphological features}

The trained CNN outputs probabilities for each predicted class for an input image.
Grad-CAM can be used to highlight the most important morphological features used for making correct and incorrect predictions, and both sets of image features are valuable for understanding what the CNN has learned.
The association between blue stellar populations at the edge of a galaxy's star-forming disk and a high-$\mathcal M$ classification, for example, strengthens our confidence in trained CNN.
Examples of the galaxy image cutout, low-$\mathcal M$ features, and high-$\mathcal M$ features are shown in Figure~\ref{fig:grad-cam}.
Below, we list the most commonly observed results.

\begin{enumerate}
\item \ion{H}{2} regions, often indicated by bright, blue, compact features, and spiral arms, usually signify that a galaxy has high gas mass fraction.
\item Red central regions due to older stellar populations tend to be associated with low $\mathcal M$.
However, red stellar populations can sometimes be conflated with dust.
\item If the flocculent outer regions of a galaxy are blue, then the CNN tends to predict that it is gas-rich, but if the outer regions are populated with redder stars (e.g., panel d of Figure~\ref{fig:grad-cam}), then the galaxy is more often predicted to be gas-poor \citep[e.g.,][]{Koopmann&Kenney04}.
\item Nearby objects in the field of view, even ones that are clearly background or foreground objects, are often considered by the CNN.
They may be highlighted as evidence for low or high $\mathcal M$ depending on their relative color to the main system; however, their contributions to the overall prediction are usually subdominant.
In Appendix~\ref{sec:perturbations}, we verify that artificial point sources are generally unimportant for the CNN's decision-making process. 
\end{enumerate}

\subsection{The value of single-band imaging}

It is clear that the CNN relies on color information to identify gas-rich or gas-poor regions.
Galaxy morphology is another useful, albeit subdominant, parameter for estimating gas mass fraction \citep[e.g.,][]{Zhang+09,Eckert+15,Teimoorinia+17}.
However, morphological features often covary with color because galaxy structures are linked to their stellar populations.
Therefore, we aim to identify the most crucial morphological features using monochromatic imaging, i.e., single-channel image cutouts with summed $g$, $r$, and $i$ flux.
We first verify that this is possible by repeating the regression analysis in Section~\ref{sec:results} using single-band imaging for \alf{}A, and recover $\mathcal M$ to within RMSE~$=0.29$~dex.
We also note that CNNs have had some success predicting other gas-phase metallicity from single-band imaging \citep[see Section 5.3 of][]{WuBoada19}.

We train a CNN to classify gas-rich and gas-poor galaxies using monochromatic imaging.
The data set, model, and optimization steps are otherwise the same as described in the previous section.
We find that the shallow CNN is able to classify monochromatic \alf{}A galaxies to over 90\% accuracy.

In Figure~\ref{fig:grad-cam}, we show Grad-CAM results on single-band imaging.
By examining the highlighted activations on monochromatic images, we are able to discern the morphological indicators of gas richness.
We find that the CNN inspects the outskirts of galaxies and highlights point source-like objects when identifying gas-rich features.
Many of these compact features are star-forming knots in spiral arms or the extended disk, but the CNN also takes background sources into consideration (e.g., panel c).
Grad-CAM also reveals that the CNN focuses on galaxy centers when identifying gas-poor features in monochromatic imaging. 
We surmise that it is relying the central surface brightness to recognize whether a galaxy is gas-poor; surprisingly, it is able to leverage this information even though no distance information is provided.

\section{Deeper imaging and future surveys} \label{sec:future}

Future optical-wavelength surveys will offer deeper imaging data sets useful for characterizing the gas properties of galaxies.
We obtain $grZ$ imaging from the DESI Legacy Imaging Surveys DR8 \citep[Legacy Survey;][]{Dey+19} in order to compare with our previous results.
Using the online interface\footnote{\url{https://legacysurvey.org/viewer}}, we query $448 \times 448$ pixel JPG cutouts at the native $0.262 \arcsec~{\rm pixel}^{-1}$ scale for both the \alf{} and xGASS samples (again using optical counterpart coordinates for the former). 
Legacy Survey imaging is deeper than that of SDSS by about two magnitudes, and has higher angular resolution (although it remains seeing-limited). 
Deep optical imaging is particularly critical for identifying low-surface brightness features in galaxies with complex star formation histories or recent gas accretion \citep[e.g.,][]{Duc+2015,Gereb+16,Hagen+16}.

Using Legacy Survey imaging, and the same training methodology as described in Section~\ref{sec:CNNs}, we find similar results for \alf{}A as before.
Our early results are promising and suggest that deeper optical imaging may be useful for improving $\mathcal M$ predictions.\footnote{It is difficult to directly compare the two imaging data sets: Legacy Survey image cutouts have an expanded field-of-view (1.96\arcmin{}) compared to SDSS imaging (1.48\arcmin{}), and the use of $Z$ rather than $i$-band imaging in the reddest channel may also affect CNN performance.}
It is also worth noting that the Legacy Survey DR8 imaging suffers from some imaging issues, such as pixel bleed, sky subtraction, and inconsistent zero-points in different bands, which may prevent the model from learning as much as it can.
These effects must be remedied if we want to maximize scientific gains through the combination of deep learning and wide-field optical/near-infrared surveys (e.g., the Vera C. Rubin Observatory (VRO) Legacy Survey of Space and Time (LSST), \citealt{LSST}; and the {\textit{Nancy Grace Roman Space Telescope} (\textit{RST}, formerly \textit{WFIRST}; \citealt{WFIRST}).

Current \HI{} surveys are mostly mass-limited, but SKA precursor surveys such as DINGO and LADUMA will be much more sensitive to gas-poor galaxy populations.
These new surveys will allow us to construct data sets similar to the xGASS representative sample or the volume-limited RESOLVE survey \citep[REsolved Spectroscopy Of a Local VolumE; e.g.,][]{Stark+16}, except with orders of magnitude more detections at the same \HI{} mass threshold.
In the future, we may be able to take deep \HI{} 21-cm line observations of some small patch of sky, and then use deep optical imaging in overlapping portions in order to generate $\mathcal M$ predictions for galaxies across the entire optical survey area \citep[e.g.,][]{TransferLearning,Khan+19}.
The methods introduced in this paper may also allow us to probe the redshift evolution of the overdensity transition regime (Section~\ref{sec:environment}) or evolution of the most relevant morphological features associated with gas richness over cosmic timescales (Section~\ref{sec:interpretation}).
These tantalizing prospects can be realized, but only if the co-evolving \HI{} and stellar mass functions \citep[e.g.,][]{Lemonias+13} and their effects on the priors baked into the trained CNN model are taken into account \citep[e.g., by sampling according to a known distribution;][]{ClassImbalance}.
Moreover, cosmic variance effects for deep \HI{} surveys need to be considered \citep[e.g.,][]{CosmicVariance}.
Finally, it is imperative to deploy robust pattern recognition algorithms in order to safeguard against out-of-distribution errors and gauge the reliability of machine learning predictions.

\section{Conclusions} \label{sec:conclusions}

In this work, we have found that deep CNNs can predict a galaxy's \HI{} mass fraction ($\mathcal M$) solely from $gri$ imaging to within RMSE~$=0.23~$dex for the \alf{}A sample, demonstrating that there is a strong connection between galaxy morphology and \HI{} content.
We have also trained a CNN for pattern recognition (PR), e.g., determining whether a galaxy is likely to be detected by an ALFALFA-like survey based on its optical imaging.
The combined regression and PR results generalize well to new test data sets, and our experiments indicate that PR threshold of $p_{\rm CNN} > 0.9$ is best-suited for optically selected galaxy samples.
We find that the CNN consistently outperforms previous machine learning methods on matched test data; using a $p_{\rm CNN} > 0.9$ PR cut, we report RMSE~$=0.20$~dex scatter for $\alpha$.100, $0.29$~dex scatter for NIBLES, and $0.22$~dex scatter for xGASS.
Our methodology can be augmented with deeper imaging or larger and more diverse galaxy samples.
With the advent of next-generation \HI{} 21-cm emission line surveys with the SKA precursor telescopes, and LSST and the \textit{Roman Space Telescope} on the horizon, it will soon be possible to generate enormous CNN-predicted \HI{} catalogs.

We are able to the probe the environmental dependence of the \HI{}-morphology relation by independently training and validating CNNs using subsamples stratified by galaxy overdensity (i.e., $\delta_5$, the normalized projected density).
For high-density environments, a CNN trained on lower-$\delta_5$ examples is unable to accurately estimate $\mathcal M$ from optical imaging.
However, if the validation set comprises galaxies in low- or intermediate-density environments, then a CNN has no trouble predicting $\mathcal M$.
We propose that in the most overdense environments, $\log(1 + \delta_5) \gtrsim 0.5$ for \alf{}A, physical processes such as ram pressure stripping, tidal interactions, and other gas depletion effects are responsible for ``breaking down'' the \HI{}-morphology relation observed in less dense environments.

We have also reformulated the problem of estimating $\mathcal M$ as a binary classification task in order to better understand how CNNs are able to distinguish gas-poor from gas-rich systems.
We use Gradient-weighted Class Activation Maps (Grad-CAM) to localize the optical features that are most important for predicting whether or not a galaxy is gas-rich.
Bright star-forming regions and clumpy blue features usually imply high $\mathcal M$, while central red bulges and older stellar populations at large radii often indicate low $\mathcal M$.
The CNN successfully distinguishes gas-rich and gas-poor galaxies with $>90\%$ accuracy using single-band optical images, implying that it is able to identify purely morphological features for estimating gas content.

We have highlighted several ways that deep learning and computer vision can be useful for understanding galaxy evolution.
Apart from predicting $\mathcal M$ and the gas fraction's reliability directly from optical imaging, CNNs can also be used to gauge the impact of co-varying galaxy properties such as environmental overdensity.
These methods are visually interpretable and provide key insights into the physical processes and stellar/ISM structures that are most closely connected to the \HI{} properties in galaxies.

\appendix

\section{Comparing CNNs to simpler models}
\label{sec:simple-models}

\begin{figure}
    \centering
    \includegraphics[width=0.97\columnwidth]{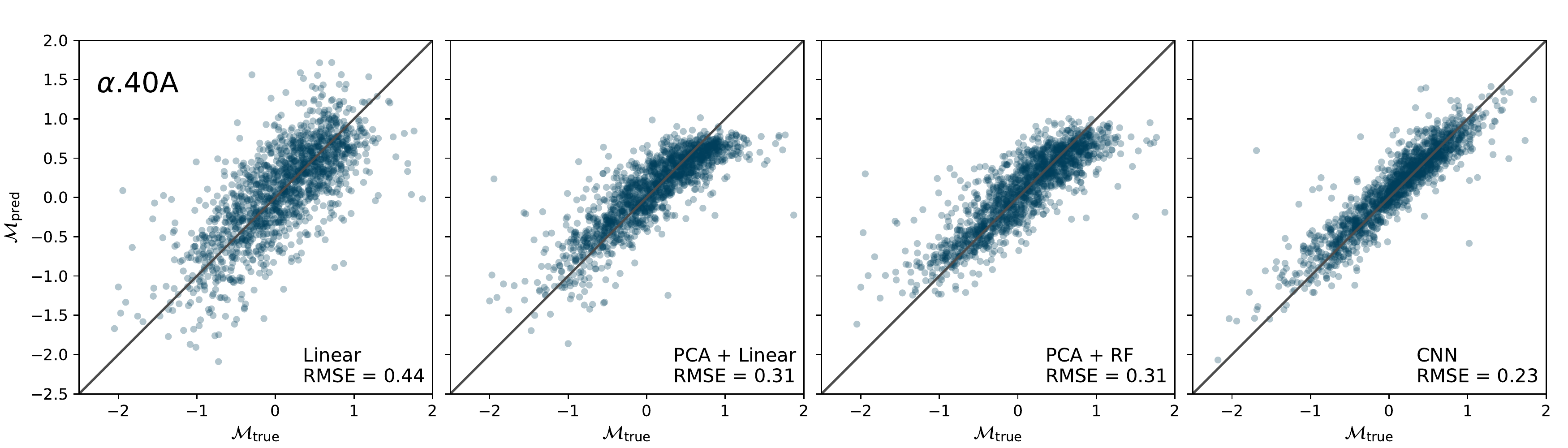}
    \includegraphics[width=0.97\columnwidth]{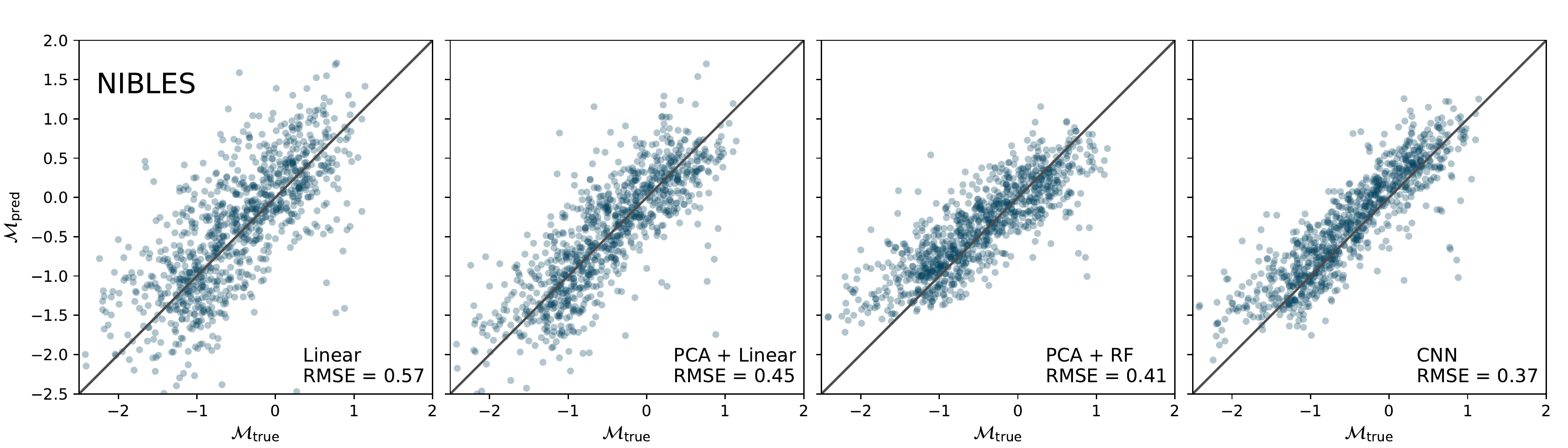}
    \caption{
    Comparison of different regression models for estimating $\mathcal M$ using the ALFALFA \alf{}A validation sample (top) and the NIBLES test sample (bottom).
    Each panel shows $\mathcal M_{\rm pred}$ plotted against $\mathcal M_{\rm true}$ for a linear model trained on block-reduced images (left), a linear model trained on PCA-processed images (center-left), a random forest model trained on the PCA-processed images (center-right), and a CNN trained on the original images (right).
    }
    \label{fig:simpler-models}
\end{figure}

Classical machine learning algorithms and statistical methods are not well-suited for operating on image data because they are not invariant to translation, scaling, or rotation.
In other words, individual pixels may represent to different galaxy features for different images, and any model that treats a pixel as a static feature will not perform well in computer vision problems.
CNNs are able to encode optimized representations of galaxy features through convolution operations, which largely do not depend on the feature's absolute location within an image.
Moreover, these invariances can be learned via redundant convolutional filters by employing data augmentation \citep[such as shifts, rotations, and crops; e.g.,][]{Dieleman+15}.

Our data set consists of SDSS image cutouts centered on the optical sources of various \HI{} catalog members.
Although the galaxies are still observed at different position angles and inclinations, and have different physical scales because they span $0 \leq z \leq 0.06$, the galaxies' central regions always occupy the center pixels.
For this reason, it may be possible to use simple models, rather than a CNN, in order to estimate the \HI{} mass fraction.
We use the \texttt{sklearn} Python package to pre-process and fit our data.

The image data are represented as arrays with $3 \times 224 \times 224$ elements.
Most simple regression models are ill-equipped to handle $150,528$ inputs at a time, so we use either one of two methods to lower the dimensionality of the input data.
The first method is to block-reduce each training image using a $7 \times 7$ kernel (also known as average binning or pooling), resulting in a $3\times 32 \times 32$-shaped input.
Each block-reduced array is then flattened into a one-dimensional vector, so that the independent variables can be written as a $N \times 3,072$ matrix, where $N$ is the number of galaxies in the training set.
The second method is to use a principal components analysis (PCA) on the training set, where only the top 16 components are flattened and saved.
The PCA-processed independent variables can be written as a $16 \times 150,528$ matrix.

After reducing the dimensionality of the training inputs, we select one of two algorithms for statistical regression.
The first algorithm is an ordinary least-squares regression to fit a low-order polynomial model.
In practice we find that a linear model always outperforms a quadratic model, so only linear regression results are included in this discussion.
The linear regression model requires 3,073 trainable parameters for block-reduced images, and 17 trainable parameters for PCA-processed images.
The second algorithm is a random forest (RF), which bootstraps (i.e., samples with replacement) 100 random decision tree estimators.
The RF regression model is optimized according to the mean squared error loss.

In Figure~\ref{fig:simpler-models}, we compare different models trained on 80\% of \alf{}A and validated on the remaining 20\% for \alf{}A (left) and for the entire NIBLES test set (right).
All model predictions are impacted by the 0.16~dex systematic offset in \HI{} mass between NIBLES and ALFALFA \citep{NIBLES}.
The linear models fit to block-reduced images do not perform well, as indicated by high scatter.
PCA-processed data provide better results than block-reduced images, although overall performance is still modest.
Ultimately, CNNs outperform all of the simpler models that we test in terms of slope and scatter.

\section{Robustness to perturbations}
\label{sec:perturbations}

\begin{figure}
    \centering
    \includegraphics[width=0.61\columnwidth]{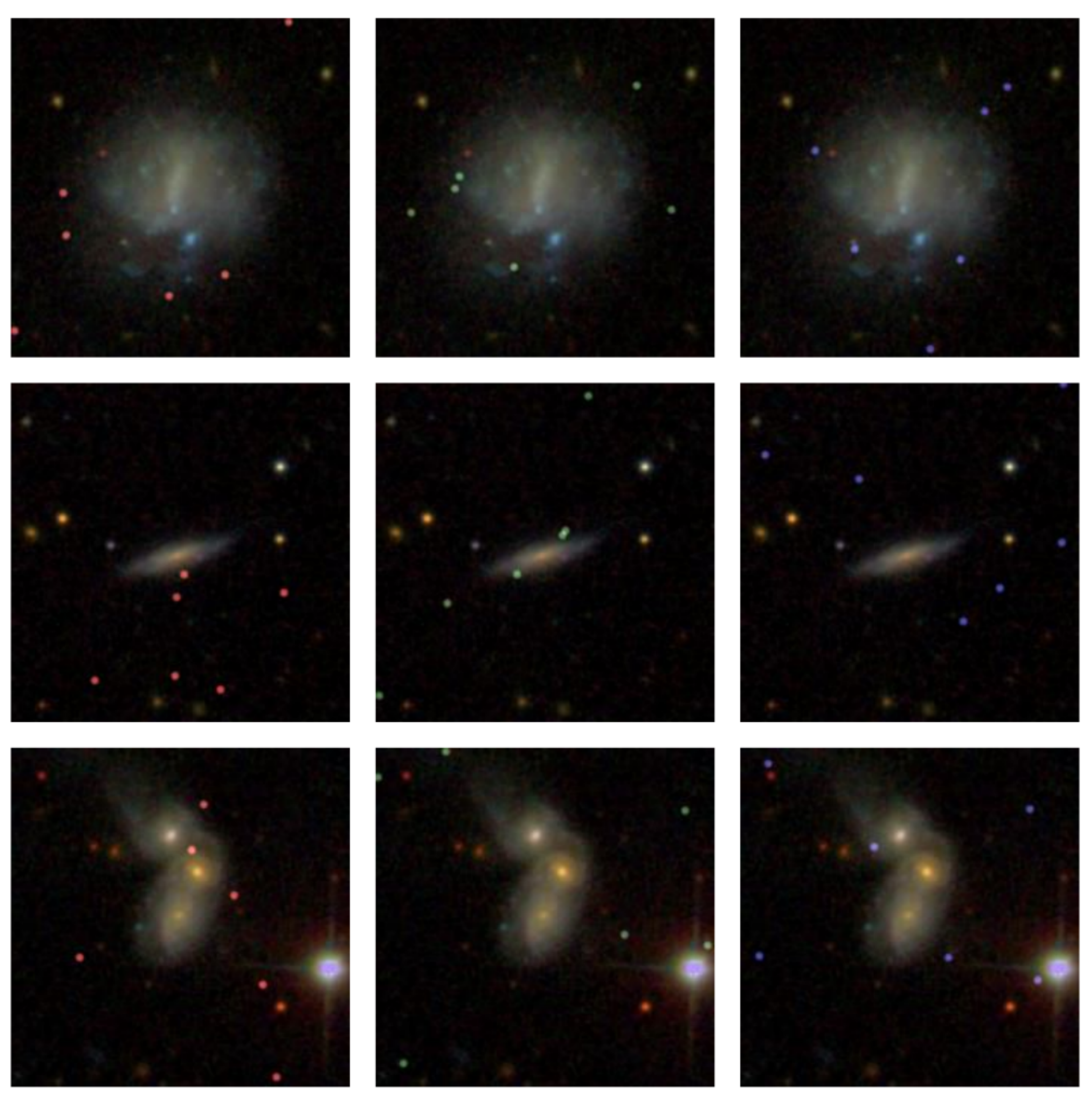}
    \includegraphics[width=0.36\columnwidth]{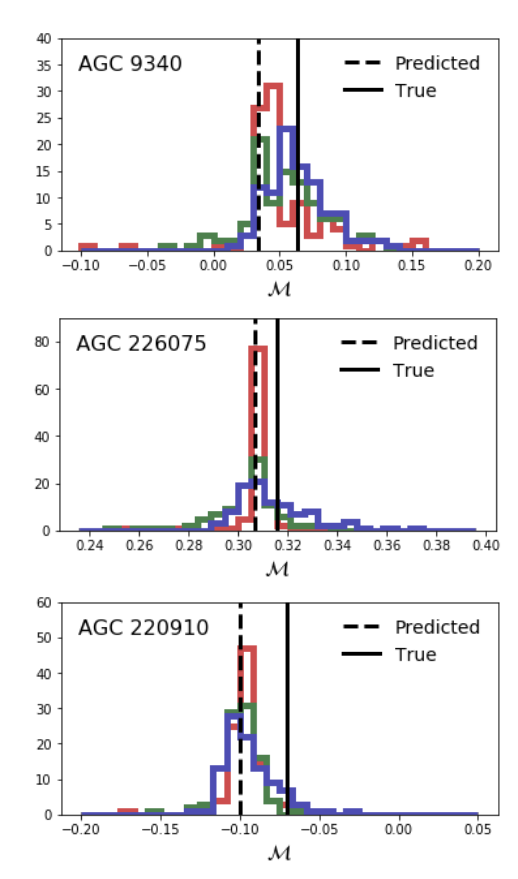}
    \caption{Examples of artificial point source injection for AGC~9340 (upper), AGC~226075 (center), and AGC~220910 (lower).
    For each row, starting from left to right, we show images with six artificial red sources, green sources, and blue sources.
    The right-most panel in each row compares the original $\mathcal M_{\rm pred}$ (dashed vertical line) and $\mathcal M_{\rm true}$ (solid vertical line) to histograms of 100 simulations for each injected source color (red, green, and blue).
    }
    \label{fig:perturbations}
\end{figure}

A concern with CNNs and deep learning algorithms is whether or not their predictions can be significantly swayed by image noise or other perturbations.
For example, galaxy images with foreground stars or other faint background sources should not cause predictions to vary wildly (unless they are located in regions where the CNN ascribes high importance, which we can probe using Grad-CAM; see Section~\ref{sec:interpretation}).
To test our method's performance when small changes are added to the images, we randomly add colored point sources (resembling artificial stars) to three representative galaxy images, and allow the CNN to infer $\mathcal M_{\rm pred}$.
Specifically, we add six artificial sources (two-dimensional circular Gaussian profiles with a 5-pixel radius, all of which are red, green, or blue) to random locations in the original image. 
Figure~\ref{fig:perturbations} shows examples of the three galaxy images with injected artificial sources.
100 trials are run for each of the three colors, and we compare these perturbed predictions to the original estimate ($\mathcal M_{\rm pred}$) and the ground truth ($\mathcal M_{\rm true}$).
We find the CNN is able to make generalized predictions that does not depend on the injected point sources; the typical scatter due to these injected sources is much smaller ($< 0.05$~dex) than the overall RMSE.
Thus, we conclude that our trained CNNs are robust to perturbations such as artificial point sources.

\bibliography{bibliography.bib} 

\begin{thebibliography}{}
\expandafter\ifx\csname natexlab\endcsname\relax\def\natexlab#1{#1}\fi

\bibitem[{{Abolfathi} {et~al.}(2018){Abolfathi}, {Aguado}, {Aguilar}, {Allende
  Prieto}, {Almeida}, {Ananna}, {Anders}, {Anderson}, {Andrews}, {Anguiano},
  {Arag{\'o}n-Salamanca}, {Argudo-Fern{\'a}ndez}, {Armengaud}, {Ata},
  {Aubourg}, {Avila-Reese}, {Badenes}, {Bailey}, {Balland}, {Barger},
  {Barrera-Ballesteros}, {Bartosz}, {Bastien}, {Bates}, {Baumgarten},
  {Bautista}, {Beaton}, {Beers}, {Belfiore}, {Bender}, {Bernardi}, {Bershady},
  {Beutler}, {Bird}, {Bizyaev}, {Blanc}, {Blanton}, {Blomqvist}, {Bolton},
  {Boquien}, {Borissova}, {Bovy}, {Bradna Diaz}, {Brandt}, {Brinkmann},
  {Brownstein}, {Bundy}, {Burgasser}, {Burtin}, {Busca}, {Ca{\~n}as},
  {Cano-D{\'\i}az}, {Cappellari}, {Carrera}, {Casey}, {Cervantes Sodi}, {Chen},
  {Cherinka}, {Chiappini}, {Choi}, {Chojnowski}, {Chuang}, {Chung}, {Clerc},
  {Cohen}, {Comerford}, {Comparat}, {Correa do Nascimento}, {da Costa},
  {Cousinou}, {Covey}, {Crane}, {Cruz-Gonzalez}, {Cunha}, {da Silva Ilha},
  {Damke}, {Darling}, {Davidson}, {Dawson}, {de Icaza Lizaola}, {de la
  Macorra}, {de la Torre}, {De Lee}, {de Sainte Agathe}, {Deconto Machado},
  {Dell'Agli}, {Delubac}, {Diamond-Stanic}, {Donor}, {Downes}, {Drory}, {du Mas
  des Bourboux}, {Duckworth}, {Dwelly}, {Dyer}, {Ebelke}, {Davis Eigenbrot},
  {Eisenstein}, {Elsworth}, {Emsellem}, {Eracleous}, {Erfanianfar},
  {Escoffier}, {Fan}, {Fern{\'a}ndez Alvar}, {Fernandez-Trincado}, {Fernand o
  Cirolini}, {Feuillet}, {Finoguenov}, {Fleming}, {Font-Ribera}, {Freischlad},
  {Frinchaboy}, {Fu}, {G{\'o}mez Maqueo Chew}, {Galbany}, {Garc{\'\i}a
  P{\'e}rez}, {Garcia-Dias}, {Garc{\'\i}a-Hern{\'a}ndez}, {Garma Oehmichen},
  {Gaulme}, {Gelfand }, {Gil-Mar{\'\i}n}, {Gillespie}, {Goddard}, {Gonz{\'a}lez
  Hern{\'a}ndez}, {Gonzalez-Perez}, {Grabowski}, {Green}, {Grier}, {Gueguen},
  {Guo}, {Guy}, {Hagen}, {Hall}, {Harding}, {Hasselquist}, {Hawley}, {Hayes},
  {Hearty}, {Hekker}, {Hernand ez}, {Hernandez Toledo}, {Hogg},
  {Holley-Bockelmann}, {Holtzman}, {Hou}, {Hsieh}, {Hunt}, {Hutchinson},
  {Hwang}, {Jimenez Angel}, {Johnson}, {Jones}, {J{\"o}nsson}, {Jullo}, {Khan},
  {Kinemuchi}, {Kirkby}, {Kirkpatrick}, {Kitaura}, {Knapp}, {Kneib},
  {Kollmeier}, {Lacerna}, {Lane}, {Lang}, {Law}, {Le Goff}, {Lee}, {Li}, {Li},
  {Lian}, {Liang}, {Lima}, {Lin}, {Long}, {Lucatello}, {Lundgren}, {Mackereth},
  {MacLeod}, {Mahadevan}, {Maia}, {Majewski}, {Manchado}, {Maraston},
  {Mariappan}, {Marques-Chaves}, {Masseron}, {Masters}, {McDermid}, {McGreer},
  {Melendez}, {Meneses-Goytia}, {Merloni}, {Merrifield}, {Meszaros}, {Meza},
  {Minchev}, {Minniti}, {Mueller}, {Muller-Sanchez}, {Muna}, {Mu{\~n}oz},
  {Myers}, {Nair}, {Nand ra}, {Ness}, {Newman}, {Nichol}, {Nidever},
  {Nitschelm}, {Noterdaeme}, {O'Connell}, {Oelkers}, {Oravetz}, {Oravetz},
  {Ort{\'\i}z}, {Osorio}, {Pace}, {Padilla}, {Palanque-Delabrouille},
  {Palicio}, {Pan}, {Pan}, {Parikh}, {P{\^a}ris}, {Park}, {Peirani},
  {Pellejero-Ibanez}, {Penny}, {Percival}, {Perez-Fournon}, {Petitjean},
  {Pieri}, {Pinsonneault}, {Pisani}, {Prada}, {Prakash}, {Queiroz}, {Raddick},
  {Raichoor}, {Barboza Rembold}, {Richstein}, {Riffel}, {Riffel}, {Rix},
  {Robin}, {Rodr{\'\i}guez Torres}, {Rom{\'a}n-Z{\'u}{\~n}iga}, {Ross},
  {Rossi}, {Ruan}, {Ruggeri}, {Ruiz}, {Salvato}, {S{\'a}nchez}, {S{\'a}nchez},
  {Sanchez Almeida}, {S{\'a}nchez-Gallego}, {Santana Rojas}, {Santiago},
  {Schiavon}, {Schimoia}, {Schlafly}, {Schlegel}, {Schneider}, {Schuster},
  {Schwope}, {Seo}, {Serenelli}, {Shen}, {Shen}, {Shetrone}, {Shull}, {Silva
  Aguirre}, {Simon}, {Skrutskie}, {Slosar}, {Smethurst}, {Smith}, {Sobeck},
  {Somers}, {Souter}, {Souto}, {Spindler}, {Stark}, {Stassun}, {Steinmetz},
  {Stello}, {Storchi-Bergmann}, {Streblyanska}, {Stringfellow}, {Su{\'a}rez},
  {Sun}, {Szigeti}, {Taghizadeh-Popp}, {Talbot}, {Tang}, {Tao}, {Tayar},
  {Tembe}, {Teske}, {Thakar}, {Thomas}, {Tissera}, {Tojeiro}, {Tremonti},
  {Troup}, {Urry}, {Valenzuela}, {van den Bosch}, {Vargas-Gonz{\'a}lez},
  {Vargas-Maga{\~n}a}, {Vazquez}, {Villanova}, {Vogt}, {Wake}, {Wang},
  {Weaver}, {Weijmans}, {Weinberg}, {Westfall}, {Whelan}, {Wilcots}, {Wild},
  {Williams}, {Wilson}, {Wood-Vasey}, {Wylezalek}, {Xiao}, {Yan}, {Yang},
  {Ybarra}, {Y{\`e}che}, {Zakamska}, {Zamora}, {Zarrouk}, {Zasowski}, {Zhang},
  {Zhao}, {Zhao}, {Zheng}, {Zheng}, {Zhou}, {Zhu}, {Zinn}, \&
  {Zou}}]{Abolfathi+18}
{Abolfathi}, B., {Aguado}, D.~S., {Aguilar}, G., {et~al.} 2018, \apjs, 235, 42

\bibitem[{{Amodei} {et~al.}(2016){Amodei}, {Olah}, {Steinhardt}, {Christiano},
  {Schulman}, \& {Man{\'e}}}]{AI-Safety}
{Amodei}, D., {Olah}, C., {Steinhardt}, J., {et~al.} 2016, arXiv e-prints,
  arXiv:1606.06565

\bibitem[{{Andrianomena} {et~al.}(2020){Andrianomena}, {Rafieferantsoa}, \&
  {Dav{\'e}}}]{Andrianomena+20}
{Andrianomena}, S., {Rafieferantsoa}, M., \& {Dav{\'e}}, R. 2020, \mnras, 492,
  5743

\bibitem[{{Barnes} {et~al.}(2001){Barnes}, {Staveley-Smith}, {de Blok},
  {Oosterloo}, {Stewart}, {Wright}, {Banks}, {Bhathal}, {Boyce}, {Calabretta},
  {Disney}, {Drinkwater}, {Ekers}, {Freeman}, {Gibson}, {Green}, {Haynes}, {te
  Lintel Hekkert}, {Henning}, {Jerjen}, {Juraszek}, {Kesteven}, {Kilborn},
  {Knezek}, {Koribalski}, {Kraan-Korteweg}, {Malin}, {Marquarding}, {Minchin},
  {Mould}, {Price}, {Putman}, {Ryder}, {Sadler}, {Schr{\"o}der}, {Stootman},
  {Webster}, {Wilson}, \& {Ye}}]{Barnes+01}
{Barnes}, D.~G., {Staveley-Smith}, L., {de Blok}, W.~J.~G., {et~al.} 2001,
  \mnras, 322, 486

\bibitem[{Bishop(1995)}]{Bishop95}
Bishop, C.~M. 1995, Neural Networks for Pattern Recognition (USA: Oxford
  University Press, Inc.)

\bibitem[{{Blanton} {et~al.}(2011){Blanton}, {Kazin}, {Muna}, {Weaver}, \&
  {Price-Whelan}}]{NSAtlas}
{Blanton}, M.~R., {Kazin}, E., {Muna}, D., {Weaver}, B.~A., \& {Price-Whelan},
  A. 2011, \aj, 142, 31

\bibitem[{{Blyth} {et~al.}(2016){Blyth}, {Baker}, {Holwerda}, {Bouchard},
  {Catinella}, {Chemin}, {Cunnama}, {Dav{\'e}}, {Faltenbacher}, {February},
  {Fern{\'a}ndez}, {Gawiser}, {Heywood}, {Kere{\v{s}}}, {Kl{\"o}ckner}, {Lah},
  {Lochner}, {Maddox}, {Makhathini}, {Moodley}, {Morganti}, {Obreschkow}, {Oh},
  {Pisano}, {Popping}, {Popping}, {Ravindranath}, {Schinnerer}, {Sheth},
  {Skelton}, {Smith}, {Srianand}, {Staveley-Smith}, {Vaccari}, {Vaisanen},
  {Walter}, {Rawlings}, {Bassett}, {Bershady}, {Briggs}, {Crawford}, {Cress},
  {Darling}, {Deane}, {de Blok}, {Elson}, {Frank}, {Henning}, {Hess}, {Hughes},
  {Jarvis}, {Kannappan}, {Katz}, {Kraan-Korteweg}, {Lehnert}, {Leroy},
  {Meurer}, {Meyer}, {Pisano}, {Schr{\"o}der}, {Smirnov}, {Somerville},
  {Stewart}, {van der Heyden}, {Verheijen}, {Wilcots}, {Williams}, {Woudt},
  {Wu}, {Zwaan}, {Zwart}, {Oosterloo}, \& {van Drie}}]{Blyth+16}
{Blyth}, S., {Baker}, A.~J., {Holwerda}, B., {et~al.} 2016, in Proceedings of
  MeerKAT Science: On the Pathway to the SKA. 25-27 May, 4

\bibitem[{{Brinchmann} {et~al.}(2004){Brinchmann}, {Charlot}, {White},
  {Tremonti}, {Kauffmann}, {Heckman}, \& {Brinkmann}}]{Brinchmann+04}
{Brinchmann}, J., {Charlot}, S., {White}, S.~D.~M., {et~al.} 2004, \mnras, 351,
  1151

\bibitem[{{Brown} {et~al.}(2017){Brown}, {Catinella}, {Cortese}, {Lagos},
  {Dav{\'e}}, {Kilborn}, {Haynes}, {Giovanelli}, \&
  {Rafieferantsoa}}]{Brown+17}
{Brown}, T., {Catinella}, B., {Cortese}, L., {et~al.} 2017, \mnras, 466, 1275

\bibitem[{{Buda} {et~al.}(2017){Buda}, {Maki}, \&
  {Mazurowski}}]{ClassImbalance}
{Buda}, M., {Maki}, A., \& {Mazurowski}, M.~A. 2017, arXiv e-prints,
  arXiv:1710.05381

\bibitem[{{Caldeira} {et~al.}(2019){Caldeira}, {Wu}, {Nord}, {Avestruz},
  {Trivedi}, \& {Story}}]{Caldeira+19}
{Caldeira}, J., {Wu}, W.~L.~K., {Nord}, B., {et~al.} 2019, Astronomy and
  Computing, 28, 100307

\bibitem[{{Catinella} {et~al.}(2010){Catinella}, {Schiminovich}, {Kauffmann},
  {Fabello}, {Wang}, {Hummels}, {Lemonias}, {Moran}, {Wu}, {Giovanelli},
  {Haynes}, {Heckman}, {Basu-Zych}, {Blanton}, {Brinchmann}, {Budav{\'a}ri},
  {Gon{\c{c}}alves}, {Johnson}, {Kennicutt}, {Madore}, {Martin}, {Rich},
  {Tacconi}, {Thilker}, {Wild}, \& {Wyder}}]{Catinella+10}
{Catinella}, B., {Schiminovich}, D., {Kauffmann}, G., {et~al.} 2010, \mnras,
  403, 683

\bibitem[{{Catinella} {et~al.}(2013){Catinella}, {Schiminovich}, {Cortese},
  {Fabello}, {Hummels}, {Moran}, {Lemonias}, {Cooper}, {Wu}, {Heckman}, \&
  {Wang}}]{Catinella+13}
{Catinella}, B., {Schiminovich}, D., {Cortese}, L., {et~al.} 2013, \mnras, 436,
  34

\bibitem[{{Catinella} {et~al.}(2018){Catinella}, {Saintonge}, {Janowiecki},
  {Cortese}, {Dav{\'e}}, {Lemonias}, {Cooper}, {Schiminovich}, {Hummels},
  {Fabello}, {Ger{\'e}b}, {Kilborn}, \& {Wang}}]{Catinella+18}
{Catinella}, B., {Saintonge}, A., {Janowiecki}, S., {et~al.} 2018, \mnras, 476,
  875

\bibitem[{{Chabrier}(2003)}]{Chabrier03}
{Chabrier}, G. 2003, \pasp, 115, 763

\bibitem[{{Chung} {et~al.}(2009){Chung}, {van Gorkom}, {Kenney}, {Crowl}, \&
  {Vollmer}}]{Chung+09}
{Chung}, A., {van Gorkom}, J.~H., {Kenney}, J. D.~P., {Crowl}, H., \&
  {Vollmer}, B. 2009, \aj, 138, 1741

\bibitem[{{{\'C}iprijanovi{\'c}} {et~al.}(2020){{\'C}iprijanovi{\'c}},
  {Snyder}, {Nord}, \& {Peek}}]{Ciprijanovic+20}
{{\'C}iprijanovi{\'c}}, A., {Snyder}, G.~F., {Nord}, B., \& {Peek}, J.~E.~G.
  2020, Astronomy and Computing, 32, 100390

\bibitem[{{Cooper} {et~al.}(2008){Cooper}, {Newman}, {Weiner}, {Yan},
  {Willmer}, {Bundy}, {Coil}, {Conselice}, {Davis}, {Faber}, {Gerke},
  {Guhathakurta}, {Koo}, \& {Noeske}}]{Cooper+08}
{Cooper}, M.~C., {Newman}, J.~A., {Weiner}, B.~J., {et~al.} 2008, \mnras, 383,
  1058

\bibitem[{{Dey} {et~al.}(2019){Dey}, {Schlegel}, {Lang}, {Blum}, {Burleigh},
  {Fan}, {Findlay}, {Finkbeiner}, {Herrera}, {Juneau}, {Landriau}, {Levi},
  {McGreer}, {Meisner}, {Myers}, {Moustakas}, {Nugent}, {Patej}, {Schlafly},
  {Walker}, {Valdes}, {Weaver}, {Y{\`e}che}, {Zou}, {Zhou}, {Abareshi},
  {Abbott}, {Abolfathi}, {Aguilera}, {Alam}, {Allen}, {Alvarez}, {Annis},
  {Ansarinejad}, {Aubert}, {Beechert}, {Bell}, {BenZvi}, {Beutler}, {Bielby},
  {Bolton}, {Brice{\~n}o}, {Buckley-Geer}, {Butler}, {Calamida}, {Carlberg},
  {Carter}, {Casas}, {Castander}, {Choi}, {Comparat}, {Cukanovaite}, {Delubac},
  {DeVries}, {Dey}, {Dhungana}, {Dickinson}, {Ding}, {Donaldson}, {Duan},
  {Duckworth}, {Eftekharzadeh}, {Eisenstein}, {Etourneau}, {Fagrelius},
  {Farihi}, {Fitzpatrick}, {Font-Ribera}, {Fulmer}, {G{\"a}nsicke},
  {Gaztanaga}, {George}, {Gerdes}, {Gontcho}, {Gorgoni}, {Green}, {Guy},
  {Harmer}, {Hernand ez}, {Honscheid}, {Huang}, {James}, {Jannuzi}, {Jiang},
  {Joyce}, {Karcher}, {Karkar}, {Kehoe}, {Kneib}, {Kueter-Young}, {Lan},
  {Lauer}, {Le Guillou}, {Le Van Suu}, {Lee}, {Lesser}, {Perreault Levasseur},
  {Li}, {Mann}, {Marshall}, {Mart{\'\i}nez-V{\'a}zquez}, {Martini}, {du Mas des
  Bourboux}, {McManus}, {Meier}, {M{\'e}nard}, {Metcalfe},
  {Mu{\~n}oz-Guti{\'e}rrez}, {Najita}, {Napier}, {Narayan}, {Newman}, {Nie},
  {Nord}, {Norman}, {Olsen}, {Paat}, {Palanque-Delabrouille}, {Peng},
  {Poppett}, {Poremba}, {Prakash}, {Rabinowitz}, {Raichoor}, {Rezaie},
  {Robertson}, {Roe}, {Ross}, {Ross}, {Rudnick}, {Safonova}, {Saha},
  {S{\'a}nchez}, {Savary}, {Schweiker}, {Scott}, {Seo}, {Shan}, {Silva},
  {Slepian}, {Soto}, {Sprayberry}, {Staten}, {Stillman}, {Stupak}, {Summers},
  {Sien Tie}, {Tirado}, {Vargas-Maga{\~n}a}, {Vivas}, {Wechsler}, {Williams},
  {Yang}, {Yang}, {Yapici}, {Zaritsky}, {Zenteno}, {Zhang}, {Zhang}, {Zhou}, \&
  {Zhou}}]{Dey+19}
{Dey}, A., {Schlegel}, D.~J., {Lang}, D., {et~al.} 2019, \aj, 157, 168

\bibitem[{{Dieleman} {et~al.}(2015){Dieleman}, {Willett}, \&
  {Dambre}}]{Dieleman+15}
{Dieleman}, S., {Willett}, K.~W., \& {Dambre}, J. 2015, \mnras, 450, 1441

\bibitem[{{Dom{\'\i}nguez S{\'a}nchez} {et~al.}(2019){Dom{\'\i}nguez
  S{\'a}nchez}, {Huertas-Company}, {Bernardi}, {Kaviraj}, {Fischer}, {Abbott},
  {Abdalla}, {Annis}, {Avila}, {Brooks}, {Buckley-Geer}, {Carnero Rosell},
  {Carrasco Kind}, {Carretero}, {Cunha}, {D'Andrea}, {da Costa}, {Davis}, {De
  Vicente}, {Doel}, {Evrard}, {Fosalba}, {Frieman}, {Garc{\'\i}a-Bellido},
  {Gaztanaga}, {Gerdes}, {Gruen}, {Gruendl}, {Gschwend}, {Gutierrez},
  {Hartley}, {Hollowood}, {Honscheid}, {Hoyle}, {James}, {Kuehn}, {Kuropatkin},
  {Lahav}, {Maia}, {March}, {Melchior}, {Menanteau}, {Miquel}, {Nord},
  {Plazas}, {Sanchez}, {Scarpine}, {Schindler}, {Schubnell}, {Smith}, {Smith},
  {Soares-Santos}, {Sobreira}, {Suchyta}, {Swanson}, {Tarle}, {Thomas},
  {Walker}, \& {Zuntz}}]{TransferLearning}
{Dom{\'\i}nguez S{\'a}nchez}, H., {Huertas-Company}, M., {Bernardi}, M.,
  {et~al.} 2019, \mnras, 484, 93

\bibitem[{{Duc} {et~al.}(2015){Duc}, {Cuillandre}, {Karabal}, {Cappellari},
  {Alatalo}, {Blitz}, {Bournaud}, {Bureau}, {Crocker}, {Davies}, {Davis}, {de
  Zeeuw}, {Emsellem}, {Khochfar}, {Krajnovi{\'c}}, {Kuntschner}, {McDermid},
  {Michel-Dansac}, {Morganti}, {Naab}, {Oosterloo}, {Paudel}, {Sarzi}, {Scott},
  {Serra}, {Weijmans}, \& {Young}}]{Duc+2015}
{Duc}, P.-A., {Cuillandre}, J.-C., {Karabal}, E., {et~al.} 2015, \mnras, 446,
  120

\bibitem[{{Eckert} {et~al.}(2015){Eckert}, {Kannappan}, {Stark}, {Moffett},
  {Norris}, {Snyder}, \& {Hoversten}}]{Eckert+15}
{Eckert}, K.~D., {Kannappan}, S.~J., {Stark}, D.~V., {et~al.} 2015, \apj, 810,
  166

\bibitem[{{Ellison} {et~al.}(2016){Ellison}, {Teimoorinia}, {Rosario}, \&
  {Mendel}}]{Ellison+16}
{Ellison}, S.~L., {Teimoorinia}, H., {Rosario}, D.~J., \& {Mendel}, J.~T. 2016,
  \mnras, 455, 370

\bibitem[{{Fabello} {et~al.}(2012){Fabello}, {Kauffmann}, {Catinella}, {Li},
  {Giovanelli}, \& {Haynes}}]{Fabello+12}
{Fabello}, S., {Kauffmann}, G., {Catinella}, B., {et~al.} 2012, \mnras, 427,
  2841

\bibitem[{{Ger{\'e}b} {et~al.}(2016){Ger{\'e}b}, {Catinella}, {Cortese},
  {Bekki}, {Moran}, \& {Schiminovich}}]{Gereb+16}
{Ger{\'e}b}, K., {Catinella}, B., {Cortese}, L., {et~al.} 2016, \mnras, 462,
  382

\bibitem[{{Giovanelli} {et~al.}(2005){Giovanelli}, {Haynes}, {Kent},
  {Perillat}, {Saintonge}, {Brosch}, {Catinella}, {Hoffman}, {Stierwalt},
  {Spekkens}, {Lerner}, {Masters}, {Momjian}, {Rosenberg}, {Springob},
  {Boselli}, {Charmand aris}, {Darling}, {Davies}, {Garcia Lambas}, {Gavazzi},
  {Giovanardi}, {Hardy}, {Hunt}, {Iovino}, {Karachentsev}, {Karachentseva},
  {Koopmann}, {Marinoni}, {Minchin}, {Muller}, {Putman}, {Pantoja}, {Salzer},
  {Scodeggio}, {Skillman}, {Solanes}, {Valotto}, {van Driel}, \& {van
  Zee}}]{Giovanelli+05}
{Giovanelli}, R., {Haynes}, M.~P., {Kent}, B.~R., {et~al.} 2005, \aj, 130, 2598

\bibitem[{{Goyal} {et~al.}(2017){Goyal}, {Doll{\'a}r}, {Girshick}, {Noordhuis},
  {Wesolowski}, {Kyrola}, {Tulloch}, {Jia}, \& {He}}]{BatchNormWeightDecay}
{Goyal}, P., {Doll{\'a}r}, P., {Girshick}, R., {et~al.} 2017, arXiv e-prints,
  arXiv:1706.02677

\bibitem[{{Hagen} {et~al.}(2016){Hagen}, {Seibert}, {Hagen}, {Nyland },
  {Neill}, {Treyer}, {Young}, {Rich}, \& {Madore}}]{Hagen+16}
{Hagen}, L. M.~Z., {Seibert}, M., {Hagen}, A., {et~al.} 2016, \apj, 826, 210

\bibitem[{{Haynes} {et~al.}(2011){Haynes}, {Giovanelli}, {Martin}, {Hess},
  {Saintonge}, {Adams}, {Hallenbeck}, {Hoffman}, {Huang}, {Kent}, {Koopmann},
  {Papastergis}, {Stierwalt}, {Balonek}, {Craig}, {Higdon}, {Kornreich},
  {Miller}, {O'Donoghue}, {Olowin}, {Rosenberg}, {Spekkens}, {Troischt}, \&
  {Wilcots}}]{Haynes+11}
{Haynes}, M.~P., {Giovanelli}, R., {Martin}, A.~M., {et~al.} 2011, \aj, 142,
  170

\bibitem[{{Haynes} {et~al.}(2018){Haynes}, {Giovanelli}, {Kent}, {Adams},
  {Balonek}, {Craig}, {Fertig}, {Finn}, {Giovanardi}, {Hallenbeck}, {Hess},
  {Hoffman}, {Huang}, {Jones}, {Koopmann}, {Kornreich}, {Leisman}, {Miller},
  {Moorman}, {O'Connor}, {O'Donoghue}, {Papastergis}, {Troischt}, {Stark}, \&
  {Xiao}}]{Haynes+18}
{Haynes}, M.~P., {Giovanelli}, R., {Kent}, B.~R., {et~al.} 2018, \apj, 861, 49

\bibitem[{{He} {et~al.}(2015){He}, {Zhang}, {Ren}, \& {Sun}}]{Resnets}
{He}, K., {Zhang}, X., {Ren}, S., \& {Sun}, J. 2015, arXiv e-prints,
  arXiv:1512.03385

\bibitem[{{He} {et~al.}(2018){He}, {Zhang}, {Zhang}, {Zhang}, {Xie}, \&
  {Li}}]{BagOfTricks}
{He}, T., {Zhang}, Z., {Zhang}, H., {et~al.} 2018, arXiv e-prints,
  arXiv:1812.01187

\bibitem[{{Hopfield}(1987)}]{Hopfield87}
{Hopfield}, J.~J. 1987, Proceedings of the National Academy of Science, 84,
  8429

\bibitem[{{Howard} \& {Gugger}(2020)}]{fastai}
{Howard}, J., \& {Gugger}, S. 2020, arXiv e-prints, arXiv:2002.04688

\bibitem[{{Huang} {et~al.}(2012){Huang}, {Haynes}, {Giovanelli}, \&
  {Brinchmann}}]{Huang+12}
{Huang}, S., {Haynes}, M.~P., {Giovanelli}, R., \& {Brinchmann}, J. 2012, \apj,
  756, 113

\bibitem[{{Huertas-Company} {et~al.}(2019){Huertas-Company}, {Rodriguez-Gomez},
  {Nelson}, {Pillepich}, {Bottrell}, {Bernardi}, {Dom{\'\i}nguez-S{\'a}nchez},
  {Genel}, {Pakmor}, {Snyder}, \& {Vogelsberger}}]{Huertas-Company+19}
{Huertas-Company}, M., {Rodriguez-Gomez}, V., {Nelson}, D., {et~al.} 2019,
  \mnras, 489, 1859

\bibitem[{{Hunter}(2007)}]{matplotlib}
{Hunter}, J.~D. 2007, Computing in Science and Engineering, 9, 90

\bibitem[{{Ivezi{\'c}} {et~al.}(2019){Ivezi{\'c}}, {Kahn}, {Tyson}, {Abel},
  {Acosta}, {Allsman}, {Alonso}, {AlSayyad}, {Anderson}, {Andrew}, \&
  et~al.}]{LSST}
{Ivezi{\'c}}, {\v{Z}}., {Kahn}, S.~M., {Tyson}, J.~A., {et~al.} 2019, \apj,
  873, 111

\bibitem[{{Jarvis} {et~al.}(2016){Jarvis}, {Taylor}, {Agudo}, {Allison},
  {Deane}, {Frank}, {Gupta}, {Heywood}, {Maddox}, {McAlpine}, {Santos},
  {Scaife}, {Vaccari}, {Zwart}, {Adams}, {Bacon}, {Baker}, {Bassett}, {Best},
  {Beswick}, {Blyth}, {Brown}, {Bruggen}, {Cluver}, {Colafrancesco}, {Cotter},
  {Cress}, {Dav{\'e}}, {Ferrari}, {Hardcastle}, {Hale}, {Harrison}, {Hatfield},
  {Klockner}, {Kolwa}, {Malefahlo}, {Marubini}, {Mauch}, {Moodley}, {Morganti},
  {Norris}, {Peters}, {Prand oni}, {Prescott}, {Oliver}, {Oozeer},
  {Rottgering}, {Seymour}, {Simpson}, {Smirnov}, \& {Smith}}]{Jarvis+16}
{Jarvis}, M., {Taylor}, R., {Agudo}, I., {et~al.} 2016, in Proceedings of
  MeerKAT Science: On the Pathway to the SKA. 25-27 May, 6

\bibitem[{{Jones} {et~al.}(2016){Jones}, {Papastergis}, {Haynes}, \&
  {Giovanelli}}]{Jones+16}
{Jones}, M.~G., {Papastergis}, E., {Haynes}, M.~P., \& {Giovanelli}, R. 2016,
  \mnras, 457, 4393

\bibitem[{{Kannappan}(2004)}]{Kannappan04}
{Kannappan}, S.~J. 2004, \apjl, 611, L89

\bibitem[{{Kauffmann} {et~al.}(2003){Kauffmann}, {Heckman}, {White}, {Charlot},
  {Tremonti}, {Brinchmann}, {Bruzual}, {Peng}, {Seibert}, {Bernardi},
  {Blanton}, {Brinkmann}, {Castander}, {Cs{\'a}bai}, {Fukugita}, {Ivezic},
  {Munn}, {Nichol}, {Padmanabhan}, {Thakar}, {Weinberg}, \&
  {York}}]{Kauffmann+03}
{Kauffmann}, G., {Heckman}, T.~M., {White}, S. D.~M., {et~al.} 2003, \mnras,
  341, 33

\bibitem[{{Khan} {et~al.}(2019){Khan}, {Huerta}, {Wang}, {Gruendl}, {Jennings},
  \& {Zheng}}]{Khan+19}
{Khan}, A., {Huerta}, E.~A., {Wang}, S., {et~al.} 2019, Physics Letters B, 795,
  248

\bibitem[{{Kinney} {et~al.}(1996){Kinney}, {Calzetti}, {Bohlin}, {McQuade},
  {Storchi-Bergmann}, \& {Schmitt}}]{Kinney+96}
{Kinney}, A.~L., {Calzetti}, D., {Bohlin}, R.~C., {et~al.} 1996, \apj, 467, 38

\bibitem[{{Koopmann} \& {Kenney}(2004)}]{Koopmann&Kenney04}
{Koopmann}, R.~A., \& {Kenney}, J. D.~P. 2004, \apj, 613, 866

\bibitem[{{Koribalski} {et~al.}(2020){Koribalski}, {Staveley-Smith},
  {Westmeier}, {Serra}, {Spekkens}, {Wong}, {Lagos}, {Obreschkow},
  {Ryan-Weber}, {Zwaan}, {Kilborn}, {Bekiaris}, {Bekki}, {Bigiel}, {Boselli},
  {Bosma}, {Catinella}, {Chauhan}, {Cluver}, {Colless}, {Courtois}, {Crain},
  {de Blok}, {D{\'e}nes}, {Duffy}, {Elagali}, {Fluke}, {For}, {Heald},
  {Henning}, {Hess}, {Holwerda}, {Howlett}, {Jarrett}, {Jones}, {Jones},
  {J{\'o}zsa}, {Jurek}, {J{\"u}tte}, {Kamphuis}, {Karachentsev}, {Kerp},
  {Keiner}, {Kraan-Korteweg}, {Lee-Waddell}, {L{\'o}pez-S{\'a}nchez}, {Madrid},
  {Meyer}, {Mould}, {Murugeshan}, {Norris}, {Oh}, {Oosterloo}, {Popping},
  {Putman}, {Reynolds}, {Rhee}, {Robotham}, {Ryder}, {Schr{\"o}der}, {Shao},
  {Stevens}, {Taylor}, {van der Hulst}, {Verdes-Montenegro}, {Wakker}, {Wang},
  {Whiting}, {Winkel}, \& {Wolf}}]{WALLABY}
{Koribalski}, B.~S., {Staveley-Smith}, L., {Westmeier}, T., {et~al.} 2020,
  arXiv e-prints, arXiv:2002.07311

\bibitem[{{Lemonias} {et~al.}(2013){Lemonias}, {Schiminovich}, {Catinella},
  {Heckman}, \& {Moran}}]{Lemonias+13}
{Lemonias}, J.~J., {Schiminovich}, D., {Catinella}, B., {Heckman}, T.~M., \&
  {Moran}, S.~M. 2013, \apj, 776, 74

\bibitem[{{Li} {et~al.}(2012){Li}, {Kauffmann}, {Fu}, {Wang}, {Catinella},
  {Fabello}, {Schiminovich}, \& {Zhang}}]{Li+12}
{Li}, C., {Kauffmann}, G., {Fu}, J., {et~al.} 2012, \mnras, 424, 1471

\bibitem[{{Liu} {et~al.}(2019){Liu}, {Jiang}, {He}, {Chen}, {Liu}, {Gao}, \&
  {Han}}]{RAdam}
{Liu}, L., {Jiang}, H., {He}, P., {et~al.} 2019, arXiv e-prints,
  arXiv:1908.03265

\bibitem[{{Loshchilov} \& {Hutter}(2017)}]{WDnotL2}
{Loshchilov}, I., \& {Hutter}, F. 2017, arXiv e-prints, arXiv:1711.05101

\bibitem[{{Lupton} {et~al.}(2004){Lupton}, {Blanton}, {Fekete}, {Hogg},
  {O'Mullane}, {Szalay}, \& {Wherry}}]{Lupton+04}
{Lupton}, R., {Blanton}, M.~R., {Fekete}, G., {et~al.} 2004, \pasp, 116, 133

\bibitem[{McKinney(2010)}]{pandas}
McKinney, W. 2010, in Proceedings of the 9th Python in Science Conference, ed.
  S.~van~der Walt \& J.~Millman, 51 -- 56

\bibitem[{{Misra}(2019)}]{Mish}
{Misra}, D. 2019, arXiv e-prints, arXiv:1908.08681

\bibitem[{{Morningstar} {et~al.}(2019){Morningstar}, {Perreault Levasseur},
  {Hezaveh}, {Blandford}, {Marshall}, {Putzky}, {Rueter}, {Wechsler}, \&
  {Welling}}]{Morningstar+19}
{Morningstar}, W.~R., {Perreault Levasseur}, L., {Hezaveh}, Y.~D., {et~al.}
  2019, \apj, 883, 14

\bibitem[{{Moster} {et~al.}(2011){Moster}, {Somerville}, {Newman}, \&
  {Rix}}]{CosmicVariance}
{Moster}, B.~P., {Somerville}, R.~S., {Newman}, J.~A., \& {Rix}, H.-W. 2011,
  \apj, 731, 113

\bibitem[{{M{\"u}ller} {et~al.}(2019){M{\"u}ller}, {Kornblith}, \&
  {Hinton}}]{LabelSmoothing}
{M{\"u}ller}, R., {Kornblith}, S., \& {Hinton}, G. 2019, arXiv e-prints,
  arXiv:1906.02629

\bibitem[{{Odekon} {et~al.}(2016){Odekon}, {Koopmann}, {Haynes}, {Finn},
  {McGowan}, {Micula}, {Reed}, {Giovanelli}, \& {Hallenbeck}}]{Odekon+16}
{Odekon}, M.~C., {Koopmann}, R.~A., {Haynes}, M.~P., {et~al.} 2016, \apj, 824,
  110

\bibitem[{{Pasquet} {et~al.}(2019){Pasquet}, {Bertin}, {Treyer}, {Arnouts}, \&
  {Fouchez}}]{Pasquet+19}
{Pasquet}, J., {Bertin}, E., {Treyer}, M., {Arnouts}, S., \& {Fouchez}, D.
  2019, \aap, 621, A26

\bibitem[{Paszke {et~al.}(2019)Paszke, Gross, Massa, Lerer, Bradbury, Chanan,
  Killeen, Lin, Gimelshein, Antiga, Desmaison, Kopf, Yang, DeVito, Raison,
  Tejani, Chilamkurthy, Steiner, Fang, Bai, \& Chintala}]{pytorch}
Paszke, A., Gross, S., Massa, F., {et~al.} 2019, in Advances in Neural
  Information Processing Systems 32, ed. H.~Wallach, H.~Larochelle,
  A.~Beygelzimer, F.~d'Alch\'{e} Buc, E.~Fox, \& R.~Garnett (Curran Associates,
  Inc.), 8024--8035

\bibitem[{Pedregosa {et~al.}(2011)Pedregosa, Varoquaux, Gramfort, Michel,
  Thirion, Grisel, Blondel, Prettenhofer, Weiss, Dubourg, Vanderplas, Passos,
  Cournapeau, Brucher, Perrot, \& Duchesnay}]{sklearn}
Pedregosa, F., Varoquaux, G., Gramfort, A., {et~al.} 2011, Journal of Machine
  Learning Research, 12, 2825

\bibitem[{{Peek} \& {Burkhart}(2019)}]{PeekBurkhart19}
{Peek}, J.~E.~G., \& {Burkhart}, B. 2019, \apjl, 882, L12

\bibitem[{{Rafieferantsoa} {et~al.}(2018){Rafieferantsoa}, {Andrianomena}, \&
  {Dav{\'e}}}]{Rafieferantsoa+18}
{Rafieferantsoa}, M., {Andrianomena}, S., \& {Dav{\'e}}, R. 2018, \mnras, 479,
  4509

\bibitem[{{Saintonge} {et~al.}(2017){Saintonge}, {Catinella}, {Tacconi},
  {Kauffmann}, {Genzel}, {Cortese}, {Dav{\'e}}, {Fletcher},
  {Graci{\'a}-Carpio}, {Kramer}, {Heckman}, {Janowiecki}, {Lutz}, {Rosario},
  {Schiminovich}, {Schuster}, {Wang}, {Wuyts}, {Borthakur}, {Lamperti}, \&
  {Roberts-Borsani}}]{Saintonge+17}
{Saintonge}, A., {Catinella}, B., {Tacconi}, L.~J., {et~al.} 2017, \apjs, 233,
  22

\bibitem[{{Salim} {et~al.}(2007){Salim}, {Rich}, {Charlot}, {Brinchmann},
  {Johnson}, {Schiminovich}, {Seibert}, {Mallery}, {Heckman}, {Forster},
  {Friedman}, {Martin}, {Morrissey}, {Neff}, {Small}, {Wyder}, {Bianchi},
  {Donas}, {Lee}, {Madore}, {Milliard}, {Szalay}, {Welsh}, \& {Yi}}]{Salim+07}
{Salim}, S., {Rich}, R.~M., {Charlot}, S., {et~al.} 2007, \apjs, 173, 267

\bibitem[{{Selvaraju} {et~al.}(2017){Selvaraju}, {Cogswell}, {Das}, {Vedantam},
  {Parikh}, \& {Batra}}]{GradCAM}
{Selvaraju}, R.~R., {Cogswell}, M., {Das}, A., {et~al.} 2017, in 2017 IEEE
  International Conference on Computer Vision (ICCV), 618--626

\bibitem[{{Serra} {et~al.}(2012){Serra}, {Oosterloo}, {Morganti}, {Alatalo},
  {Blitz}, {Bois}, {Bournaud}, {Bureau}, {Cappellari}, {Crocker}, {Davies},
  {Davis}, {de Zeeuw}, {Duc}, {Emsellem}, {Khochfar}, {Krajnovi{\'c}},
  {Kuntschner}, {Lablanche}, {McDermid}, {Naab}, {Sarzi}, {Scott}, {Trager},
  {Weijmans}, \& {Young}}]{Serra+12}
{Serra}, P., {Oosterloo}, T., {Morganti}, R., {et~al.} 2012, \mnras, 422, 1835

\bibitem[{{Serra} {et~al.}(2015){Serra}, {Koribalski}, {Kilborn}, {Allison},
  {Amy}, {Ball}, {Bannister}, {Bell}, {Bock}, {Bolton}, {Bowen}, {Boyle},
  {Broadhurst}, {Brodrick}, {Brothers}, {Bunton}, {Chapman}, {Cheng},
  {Chippendale}, {Chung}, {Cooray}, {Cornwell}, {DeBoer}, {Diamond}, {Forsyth},
  {Gough}, {Gupta}, {Hampson}, {Harvey-Smith}, {Hay}, {Hayman}, {Heywood},
  {Hotan}, {Hoyle}, {Humphreys}, {Indermuehle}, {Jacka}, {Jackson}, {Jackson},
  {Jeganathan}, {Johnston}, {Joseph}, {Kamphuis}, {Leach}, {Lenc}, {Lensson},
  {Mackay}, {Marquarding}, {Marvil}, {McClure-Griffiths}, {McConnell}, {Meyer},
  {Mirtschin}, {Neuhold}, {Ng}, {Norris}, {O'Sullivan}, {Pathikulangara},
  {Pearce}, {Phillips}, {Popping}, {Qiao}, {Reynolds}, {Roberts}, {Sault},
  {Schinckel}, {Shaw}, {Shimwell}, {Staveley-Smith}, {Storey}, {Sweetnam},
  {Troup}, {Tzioumis}, {Voronkov}, {Westmeier}, {Whiting}, {Wilson}, {Wong}, \&
  {Wu}}]{Serra+15}
{Serra}, P., {Koribalski}, B., {Kilborn}, V., {et~al.} 2015, \mnras, 452, 2680

\bibitem[{{Simonyan} {et~al.}(2013){Simonyan}, {Vedaldi}, \&
  {Zisserman}}]{Saliency}
{Simonyan}, K., {Vedaldi}, A., \& {Zisserman}, A. 2013, arXiv e-prints,
  arXiv:1312.6034

\bibitem[{{Smith}(2018)}]{OneCycle}
{Smith}, L.~N. 2018, arXiv e-prints, arXiv:1803.09820

\bibitem[{{Spergel} {et~al.}(2015){Spergel}, {Gehrels}, {Baltay}, {Bennett},
  {Breckinridge}, {Donahue}, {Dressler}, {Gaudi}, {Greene}, {Guyon}, {Hirata},
  {Kalirai}, {Kasdin}, {Macintosh}, {Moos}, {Perlmutter}, {Postman},
  {Rauscher}, {Rhodes}, {Wang}, {Weinberg}, {Benford}, {Hudson}, {Jeong},
  {Mellier}, {Traub}, {Yamada}, {Capak}, {Colbert}, {Masters}, {Penny},
  {Savransky}, {Stern}, {Zimmerman}, {Barry}, {Bartusek}, {Carpenter}, {Cheng},
  {Content}, {Dekens}, {Demers}, {Grady}, {Jackson}, {Kuan}, {Kruk}, {Melton},
  {Nemati}, {Parvin}, {Poberezhskiy}, {Peddie}, {Ruffa}, {Wallace}, {Whipple},
  {Wollack}, \& {Zhao}}]{WFIRST}
{Spergel}, D., {Gehrels}, N., {Baltay}, C., {et~al.} 2015, arXiv e-prints,
  arXiv:1503.03757

\bibitem[{{Stark} {et~al.}(2016){Stark}, {Kannappan}, {Eckert}, {Florez},
  {Hall}, {Watson}, {Hoversten}, {Burchett}, {Guynn}, {Baker}, {Moffett},
  {Berlind}, {Norris}, {Haynes}, {Giovanelli}, {Leroy}, {Pisano}, {Wei},
  {Gonzalez}, \& {Calderon}}]{Stark+16}
{Stark}, D.~V., {Kannappan}, S.~J., {Eckert}, K.~D., {et~al.} 2016, \apj, 832,
  126

\bibitem[{{Stevens} {et~al.}(2019){Stevens}, {Diemer}, {Lagos}, {Nelson},
  {Pillepich}, {Brown}, {Catinella}, {Hernquist}, {Weinberger}, {Vogelsberger},
  \& {Marinacci}}]{Stevens+19}
{Stevens}, A. R.~H., {Diemer}, B., {Lagos}, C. d.~P., {et~al.} 2019, \mnras,
  483, 5334

\bibitem[{{Teimoorinia} {et~al.}(2017){Teimoorinia}, {Ellison}, \&
  {Patton}}]{Teimoorinia+17}
{Teimoorinia}, H., {Ellison}, S.~L., \& {Patton}, D.~R. 2017, \mnras, 464, 3796

\bibitem[{{Tremonti} {et~al.}(2004){Tremonti}, {Heckman}, {Kauffmann},
  {Brinchmann}, {Charlot}, {White}, {Seibert}, {Peng}, {Schlegel}, {Uomoto},
  {Fukugita}, \& {Brinkmann}}]{Tremonti+04}
{Tremonti}, C.~A., {Heckman}, T.~M., {Kauffmann}, G., {et~al.} 2004, \apj, 613,
  898

\bibitem[{{van der Walt} {et~al.}(2011){van der Walt}, {Colbert}, \&
  {Varoquaux}}]{numpy}
{van der Walt}, S., {Colbert}, S.~C., \& {Varoquaux}, G. 2011, Computing in
  Science and Engineering, 13, 22

\bibitem[{{van Driel} {et~al.}(2016){van Driel}, {Butcher}, {Schneider},
  {Lehnert}, {Minchin}, {Blyth}, {Chemin}, {Hallet}, {Joseph}, {Kotze},
  {Kraan-Korteweg}, {Olofsson}, \& {Ramatsoku}}]{NIBLES}
{van Driel}, W., {Butcher}, Z., {Schneider}, S., {et~al.} 2016, \aap, 595, A118

\bibitem[{{Virtanen} {et~al.}(2019){Virtanen}, {Gommers}, {Oliphant},
  {Haberland}, {Reddy}, {Cournapeau}, {Burovski}, {Peterson}, {Weckesser},
  {Bright}, {van der Walt}, {Brett}, {Wilson}, {Jarrod Millman}, {Mayorov},
  {Nelson}, {Jones}, {Kern}, {Larson}, {Carey}, {Polat}, {Feng}, {Moore}, {Vand
  erPlas}, {Laxalde}, {Perktold}, {Cimrman}, {Henriksen}, {Quintero}, {Harris},
  {Archibald}, {Ribeiro}, {Pedregosa}, {van Mulbregt}, \&
  {Contributors}}]{scipy}
{Virtanen}, P., {Gommers}, R., {Oliphant}, T.~E., {et~al.} 2019, arXiv
  e-prints, arXiv:1907.10121

\bibitem[{{Wu} \& {Boada}(2019)}]{WuBoada19}
{Wu}, J.~F., \& {Boada}, S. 2019, \mnras, 484, 4683

\bibitem[{{Zeiler} \& {Fergus}(2013)}]{VisualizingCNNs}
{Zeiler}, M.~D., \& {Fergus}, R. 2013, arXiv e-prints, arXiv:1311.2901

\bibitem[{{Zhang} {et~al.}(2018){Zhang}, {Goodfellow}, {Metaxas}, \&
  {Odena}}]{SimpleSelfAttention}
{Zhang}, H., {Goodfellow}, I., {Metaxas}, D., \& {Odena}, A. 2018, arXiv
  e-prints, arXiv:1805.08318

\bibitem[{{Zhang} {et~al.}(2019){Zhang}, {Lucas}, {Hinton}, \&
  {Ba}}]{LookAhead}
{Zhang}, M.~R., {Lucas}, J., {Hinton}, G., \& {Ba}, J. 2019, arXiv e-prints,
  arXiv:1907.08610

\bibitem[{{Zhang} {et~al.}(2009){Zhang}, {Li}, {Kauffmann}, {Zou}, {Catinella},
  {Shen}, {Guo}, \& {Chang}}]{Zhang+09}
{Zhang}, W., {Li}, C., {Kauffmann}, G., {et~al.} 2009, \mnras, 397, 1243

\bibitem[{{Zhou} {et~al.}(2016){Zhou}, {Khosla}, {Lapedriza}, {Oliva}, \&
  {Torralba}}]{CAM}
{Zhou}, B., {Khosla}, A., {Lapedriza}, A., {Oliva}, A., \& {Torralba}, A. 2016,
  in 2016 IEEE Conference on Computer Vision and Pattern Recognition (CVPR),
  2921--2929

\bibitem[{{Zhu} {et~al.}(2014){Zhu}, {Berndsen}, {Madsen}, {Tan}, {Stairs},
  {Brazier}, {Lazarus}, {Lynch}, {Scholz}, {Stovall}, {Ransom}, {Banaszak},
  {Biwer}, {Cohen}, {Dartez}, {Flanigan}, {Lunsford}, {Martinez}, {Mata},
  {Rohr}, {Walker}, {Allen}, {Bhat}, {Bogdanov}, {Camilo}, {Chatterjee},
  {Cordes}, {Crawford}, {Deneva}, {Desvignes}, {Ferdman}, {Freire}, {Hessels},
  {Jenet}, {Kaplan}, {Kaspi}, {Knispel}, {Lee}, {van Leeuwen}, {Lyne},
  {McLaughlin}, {Siemens}, {Spitler}, \& {Venkataraman}}]{Zhu+14}
{Zhu}, W.~W., {Berndsen}, A., {Madsen}, E.~C., {et~al.} 2014, \apj, 781, 117

\end{thebibliography}

\software{Numpy \citep{numpy}, scikit-learn \citep{sklearn}, Scipy \citep{scipy}, matplotlib \citep{matplotlib}, Pandas \citep{pandas}, Pytorch \citep{pytorch}, Fastai (\url{https://github.com/fastai/fastai})}

\acknowledgments
The author would like to thank the anonymous referee for useful and detailed comments that have significantly improved this manuscript.
The author would like to thank Josh Peek for suggesting the idea of using CNNs to probe galaxy environments and many other useful discussions.
The author also thanks Luke Leisman and Mike Jones for helpful conversations regarding the ALFALFA data.
The author acknowledges support from the National Science Foundation under grants NSF AST-1517908 and NSF AST-1616177, and also thanks the Pascal Institute for their hospitality (The Self-organised Star Formation Process program).
This research was supported by the Munich Institute for Astro- and Particle Physics (MIAPP) which is funded by the Deutsche Forschungsgemeinschaft (DFG, German Research Foundation) under Germany's Excellence Strategy - EXC-2094 - 390783311.
This work made use of Google Colab and Google Compute Engine.

Funding for the Sloan Digital Sky Survey IV has been provided by the Alfred P. Sloan Foundation, the U.S. Department of Energy Office of Science, and the Participating Institutions. SDSS-IV acknowledges
support and resources from the Center for High-Performance Computing at
the University of Utah. The SDSS web site is www.sdss.org.

SDSS-IV is managed by the Astrophysical Research Consortium for the 
Participating Institutions of the SDSS Collaboration including the 
Brazilian Participation Group, the Carnegie Institution for Science, 
Carnegie Mellon University, the Chilean Participation Group, the French Participation Group, Harvard-Smithsonian Center for Astrophysics, 
Instituto de Astrof\'isica de Canarias, The Johns Hopkins University, Kavli Institute for the Physics and Mathematics of the Universe (IPMU) / 
University of Tokyo, the Korean Participation Group, Lawrence Berkeley National Laboratory, 
Leibniz Institut f\"ur Astrophysik Potsdam (AIP),  
Max-Planck-Institut f\"ur Astronomie (MPIA Heidelberg), 
Max-Planck-Institut f\"ur Astrophysik (MPA Garching), 
Max-Planck-Institut f\"ur Extraterrestrische Physik (MPE), 
National Astronomical Observatories of China, New Mexico State University, 
New York University, University of Notre Dame, 
Observat\'ario Nacional / MCTI, The Ohio State University, 
Pennsylvania State University, Shanghai Astronomical Observatory, 
United Kingdom Participation Group,
Universidad Nacional Aut\'onoma de M\'exico, University of Arizona, 
University of Colorado Boulder, University of Oxford, University of Portsmouth, 
University of Utah, University of Virginia, University of Washington, University of Wisconsin, 
Vanderbilt University, and Yale University.
\end{document}